\documentclass[preprint,showpacs,preprintnumbers,amsmath,amssymb,dvipdfmx]{revtex4}
\usepackage{graphicx}
\usepackage{dcolumn}
\usepackage{bm}
\usepackage{amsmath} 
\usepackage{color} 
\def\v#1{\mbox{\boldmath $#1$}}

\begin{document}

\title{Thermoosmosis of a near-critical binary  fluid mixture: \\
a general formulation  and universal flow direction}

\author{Youhei Fujitani}
 \email{youhei@appi.keio.ac.jp}
\affiliation{School of Fundamental Science and Technology,
Keio University, 
Yokohama 223-8522, Japan}

\author{Shunsuke Yabunaka}
\email{yabunaka123@gmail.com}
\affiliation{Advanced Science Research Center, Japan Atomic Energy Agency, Tokai, 319-1195, Japan}


\date{\today}

\begin{abstract}
We consider 
{a} binary fluid mixture, which lies in the one-phase region
near the demixing critical point, {and study} its transport through a capillary tube 
{linking two large reservoirs}.
We assume that short-range interactions cause 
preferential adsorption of one component  on the tube's wall. The adsorption layer can   
{become much thicker than} the molecular size, which enables us to apply 
{hydrodynamics based on a coarse-grained free-energy} functional.
For linear transport phenomena 
induced by gradients of the pressure, composition, and temperature along a cylindrical tube,   
we obtain the formulas of the Onsager coefficients to extend 
{our previous results on isothermal transport,}
assuming the critical composition in the middle of each reservoir 
 in the reference equilibrium state.
{Among the linear transport phenomena},
{we focus on thermoosmosis --- mass flow due to a temperature gradient}. We
{explicitly derive a formula for the thermal force density, which  is nonvanishing in the adsorption layer
and causes thermoosmosis.}  This formula for a near-critical binary fluid
mixture is an extension of the conventional formula for a one-component fluid,
expressed in terms of local excess enthalpy.
We predict that
{the direction of thermoosmotic flow} of a mixture near the upper (lower) consolute point is the same as (opposite to) 
that of the temperature gradient, {irrespective of which component is adsorbed on the wall.}
Our procedure would also be applied to 
dynamics of a soft material, whose mesoscopic inhomogeneity 
can be described by a coarse-grained free-energy functional.
\end{abstract}
\maketitle
\section{\label{sec:intro}Introduction}
{Osmotic transport of a fluid through a channel at micrometer, or smaller scales, has
gained much attention because it is applied in lab-on-a-chip processes \cite{breg, lee, shin, shakib, chen} and
involves fundamental problems in nonequilibrium physics \cite{piazza, Wurger, marbach,  ganti, mang, frenkel}.
A gradient of temperature (concentration)} {along the channel induces a} flow, 
{called thermoosmosis (diffusioosmosis),
because of force density generated in a heterogeneous layer near the wall \cite{derja,derja2,anders}};
thermoosmosis {does not involve the buoyancy} responsible for 
thermal convection.
Derjaguin and his coworkers rationalize thermoosmosis and diffusioosmosis
in terms of the continuum description
\cite{derja3, derja4, derja5}.  \\

\noindent
For a one-component fluid, 
{Derjaguin and Sidorenkov (DS) \cite{derja3} derived a formula} 
{of the thermal force density}, {which causes thermoosmosis}.
 {According to DS's formula,
the direction of the flow
is the same as (opposite to) that of the temperature gradient 
if the excess enthalpy density is negative (positive) near the wall.  
This is} naively expected since the flow in this direction tends to eliminate the temperature gradient  
by carrying the fluid with lower (higher) enthalpy to the region with higher (lower) temperature. 
However, there {exist several difficulties when applying this formula to experimental systems as follows.} 
{Firstly, the thickness of the heterogeneous layer is microscopic
and the continuum description cannot incorporate details inside the layer very precisely \cite{ganti, frenkel}}.
Secondly, the local excess enthalpy is
not easy to access experimentally. Its 
evaluation based on microscopic models has been studied {\cite{fu,ganti}, whereas 
well-definedness of its microscopic expression is questioned especially near the surface \cite{ganti,Anzini}. 
{In Ref.~\cite{ganti}, the authors extend DS's formula for a multicomponent fluid}
{in  the continuum description, while noticing the difficulties  discussed above.}
Therefore, it remains often hard to understand even the flow direction in thermoosmosis \cite{piazza}.
 \\

\begin{figure}
\includegraphics[width=8cm]{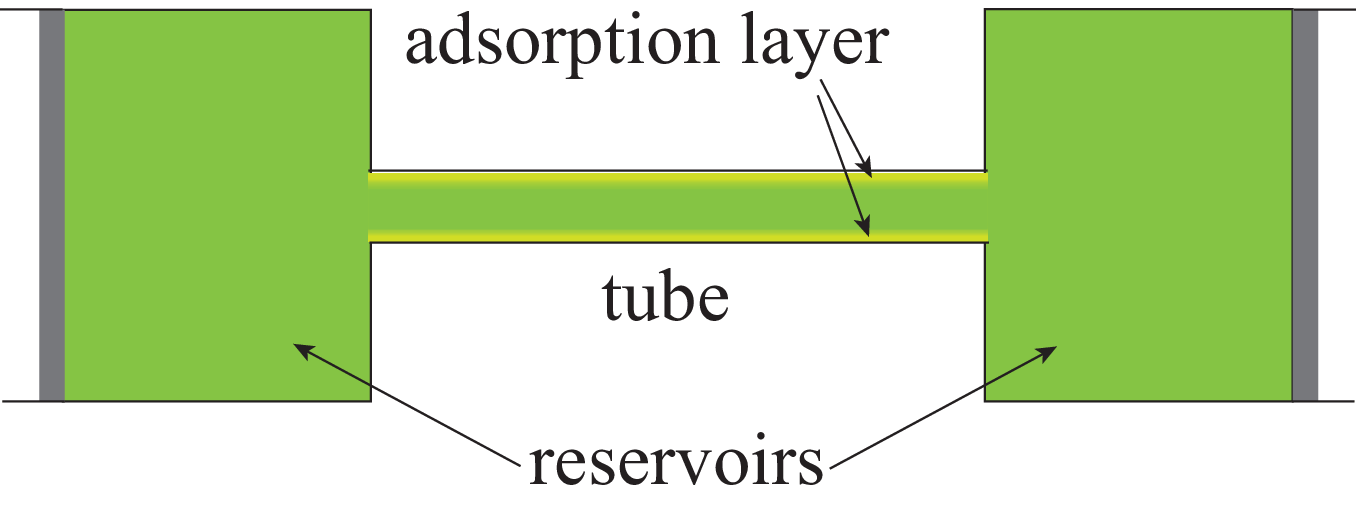}
\caption{Schematic of a situation supposed in our formulation.
A mixture is filled in the container composed of two reservoirs
and a capillary tube connecting them.  One component drawn in yellow is preferentially adsorbed
by the tube's wall, {which is impermeable and adiabatic.}
There may be preferential adsorption onto the reservoir's wall, which is not supposed in this figure. 
Thick walls represent pistons.  Imposing difference in pressure, composition, and/or temperature  
between the reservoirs generates a mass flow through the tube.
}
\label{fig:tube}
\end{figure}

\noindent
{Thermoosmosis has not been studied in relation to critical phenomena to our best knowledge.}
{In the present study, for a binary fluid mixture lying in the one-phase region
close to the demixing critical point, we extend our previous study on isothermal dynamics \cite{pipe}
to cover non-isothermal dynamics of thermoosmosis.   Below, this mixture, simply referred to as a mixture,} 
is assumed to be filled in a container composed of two large reservoirs and {a capillary tube connecting them. The tube's wall}
{is impermeable and adiabatic.  
Differences} in temperature, composition, and pressure 
can be imposed between the two reservoirs (Fig.~\ref{fig:tube}). 
 A short-range interaction is assumed between each mixture component and tube's wall, which in general
attracts one component more than the other. 
The resulting preferential adsorption (PA),
{becoming remarkable owing to large osmotic susceptibility \cite{beysens1982,schol}, {has been studied} 
for a long time \cite{binder, diehl86, diehl97, law}.}
{The adsorption layer can be much thicker than} the molecular size.
Thus, it is expected that we can study
transport of a mixture through a tube in terms of the continuum description,
avoiding its difficulty mentioned in the last paragraph.   
In Ref.~\cite{pipe}, applying the hydrodynamics
based on a coarse-grained free-energy functional \cite{yabufuji}, 
the present authors studied isothermal transport in the linear regime with respect to 
thermodynamic forces to calculate the involved Onsager coefficients
and conductance in diffusiooosmosis.   
\\

\noindent
Order-parameter fluctuations in a mixture,
being significant on length scales smaller than the correlation length, enhance
the transport coefficients to cause universal properties \cite{kawasaki,sighalhoh, sengers, folkmos, onukibook}.
However, the PA keeps the mixture composition in a tube away from the critical one, in particular near the tube's wall, 
and thus the critical enhancement does not affect diffusioosmosis of a mixture significantly \cite{pipe}.
Hydrodynamics can be applied to a flow in the tube, where the correlation length
is locally smaller than a typical length of the flow.
{It is also suggested in Ref.~\cite{pipe} that, in a critical regime}, the mixture velocity due to diffusioosmosis
far from a flat surface 
should exhibit a power-law dependence {on the difference between the mixture temperature and the critical temperature} 
{if the adsorption is sufficiently strong.  This   
originates from the universal order-parameter profile at equilibrium} \cite{fisher-degennes, RJ}. 
The same power law is numerically suggested for the
diffusiophoretic mobility of a colloidal particle in a mixture \cite{diffphore, diffdrop}.
\\

\noindent
In the present study, we extend our previous procedure of Ref.~\cite{pipe} 
to cover non-isothermal dynamics.
Our general formulation is described in Sections \ref{sec:thermo} and \ref{sec:hydro}\@.
We employ the hydrodynamics under inhomogeneous temperature
formulated from a coarse-grained free-energy functional \cite{dvw,gonn}.
{Imposing the no-slip condition at the tube's wall and neglecting effects of the} {tube's
edges}, 
we discuss flow fields in the tube in the linear regime in Section \ref{sec:tubefield}\@.
In {Sections \ref{sec:formulae} and \ref{sec:thermoosmo}}, we derive  
formulas for the Onsager coefficients and a formula 
of the thermal force density for a cylindrical tube, assuming the total mass density
to be homogeneous inside the tube and the mixture composition {to be critical in the middle of 
a reservoir} in the reference equilibrium state.
The former formulas include
extensions of our previous results of Ref.~\cite{pipe} to non-isothermal transport, whereas the latter
can be regarded as an extension of DS's formula to a mixture considered here.  
We apply the renormalized local functional theory \cite{fisher-degennes,rlft}
to specify the free-energy functional in Section \ref{sec:free}, and rewrite our formulas  
in Section \ref{sec:eval}\@.   {This theory
 can incorporate the effects of the critical fluctuations} {when the correlation length is spatially 
varying as is the case in the adsorption layer.}
In Section \ref{sec:num}, we focus on thermoosmosis to show numerical results,
and predict that, {irrespective of which component is adsorbed on the wall},
the flow direction is the same as (opposite to) the direction of the temperature gradient  in thermoososis of
a mixture near the upper (lower) consolute point.
{Further discussion and summary are given in Section \ref{sec:disc}\@}.
Some of our results are also discussed in the companion Letter \cite{complett}.

\section{Formulation\label{sec:form}}
We write 
$\rho_{\rm a}$ ($\rho_{\rm b}$) for the mass density of a mixture component named a (b). 
The sum $\rho_{\rm a}+\rho_{\rm b}$ is denoted by $\rho$, which represents
the total mass density, whereas the difference $\rho_{\rm a}-\rho_{\rm b}$ is denoted by $\varphi$.
We write $c_n$ for {the mass fraction} $\rho_n/\rho$ and
$\mu_{n}$ for the chemical potential conjugate to $\rho_{n}$, where $n$ is a or b.
In an equilibrium mixture with homogeneous mass densities, $\mu_{n}$ is a function of the
temperature (denoted by $T$), pressure ($P$), and $c_{\rm a}$, {and is also a function of
$T$, $\rho$, and $\varphi$}.
We write $\mu_\pm$ for  $\left(\mu_{\rm a}\pm\mu_{\rm b}\right)/2$;
$\rho$ and $\varphi$ are conjugate to $\mu_+$ and $\mu_-$, respectively.  
Difference of a quantity in the left reservoir  subtracted from
the quantity in the right is indicated by $\delta$.
For example, $\delta \mu_{\rm a}$ denotes the difference in $\mu_{\rm a}$ between the
reservoirs.
If the reservoir's wall adsorbs a component more, $\varphi$ can be inhomogeneous near the wall.
For such a quantity, $\delta$ indicates
the difference between the central regions of the reservoirs.
The difference between the pressures on the pistons is given by $\delta P$.

\subsection{Thermodynamics\label{sec:thermo}}
We first consider entropy fluctuations of an equilibrium mixture
in the isolated container with the pistons fixed (Fig.~\ref{fig:tube}).
The total mass of the component ${n}$ in the right reservoir is denoted by ${\cal M}_{{n} {\rm R}}$. 
Neglecting the contribution from the mixture in the tube,
we can regard the total entropy of the mixture in the container, denoted by $S$, as a function of
${\cal M}_{{\rm aR}}$, ${\cal M}_{{\rm bR}}$, 
and the internal energy of the mixture in the right reservoir, ${\cal U}_{\rm R}$.
With $t$ denoting the time, \begin{equation}
 \frac{dS}{dt}=
\frac{d{\cal U}_{\rm R}}{dt}\delta\left(\frac{1}{T}\right)+
\frac{d{\cal M}_{{n}{\rm R}}}{dt}\delta\left(\frac{-\mu_n}{T}\right)
\label{eqn:DeltaS}\end{equation}
holds up to the second order of the magnitudes of the deviations.
Repeated indices are summed hereafter.
Equation (\ref{eqn:DeltaS}) is included in Eq.~(XV-55) of Ref.~\cite{gromaz}.
The thermodynamic fluxes are given by the time derivatives on the right-hand side (RHS)
and are driven by the conjugate thermodynamic forces, $\delta(1/T)$ 
and $-\delta(\mu_n/T)$.  They are respectively
the partial derivatives
of $S$ with respect to ${\cal U}_{\rm R}$ and ${\cal M}_{n{\rm R}}$ \cite{engineer}. \\

\noindent
We assume that weak thermodynamic forces are imposed on an
equilibrium mixture.  This equilibrium state is close to the critical point and is referred to as reference state.
We add a superscript $^{({\rm ref})}$ to a quantity in the middle of a reservoir in the reference state.
For example, {$\rho_n^{({\rm ref})}$ denotes $\rho_n$} in the central regions in the reference state.
Writing $u$ for the internal energy per unit volume, we
apply the Gibbs-Duhem {(GD)} relation to obtain  
\begin{equation}
\left( \begin{array}{c} -\delta \left(P/T\right)\\ -\delta \left(\mu_-/T\right)\\ \delta \left(1/T\right) \end{array}\right)
=\left( \begin{array}{ccc}\rho_{\rm a}^{({\rm ref})}& \rho_{\rm b}^{({\rm ref})}& u^{({\rm ref})}\\ 1/2& -1/2&0\\ 0&0&1\end{array}\right)
\left(\begin{array}{c} -\delta \left(\mu_{\rm a}/T\right)\\-\delta\left(\mu_{\rm b}/T\right)\\ \delta\left(1/T\right)\end{array}\right)
\ ,\label{eqn:newforce}\end{equation} whose left-hand side (LHS) gives a new set of thermodynamic forces.  
We write $\Theta$ for the coefficient matrix on the RHS above.
The thermodynamic fluxes conjugate to the respective components of the new set, denoted by
${\cal I}$, ${\cal J}$, and ${\cal K}$, are defined by the first equality of
\begin{equation}
\left(\Theta^{-1}\right)^{\rm T} 
 \left(\begin{array}{c} d{\cal M}_{\rm aR}/(dt)\\  d{\cal M}_{\rm bR}/(dt)\\ d{\cal U}_{\rm R}/(dt) \end{array}\right)
=\left( \begin{array}{c} {\cal I}\\ {\cal J}\\{\cal K}\end{array}\right)=L
\left( \begin{array}{c} -\delta \left(P/T\right)\\ -\delta \left(\mu_-/T\right)\\ \delta \left(1/T\right)\end{array}\right)
\ ,\label{eqn:Theta}\end{equation}
where the superscript $^{\rm T}$
indicates the transposition.  The second equality above represents linear phenomenological equations;
a matrix $L$ is here introduced and their components are Onsager coefficients considered {later}.
\\

\noindent
The partial volume and partial entropy per unit mass of the component $n$, are denoted by 
${\bar v}_n$ and ${\bar s}_n$, respectively.  {In an equilibrium mixture with homogeneous densities}, we have
\begin{equation}
{\bar v}_n=\left.\frac{\partial \mu_n}{\partial P}\right)_{T, c_{\rm a}}\quad{\rm and}\quad
{\bar s}_n=-\left.\frac{\partial \mu_n}{\partial T}\right)_{P, c_{\rm a}}
\ ,\label{eqn:partials}\end{equation}
where the subscript of a right parenthesis indicates the fixed variables in the partial differentiation.
Writing $s$ for the entropy per unit volume, we have
\begin{equation}
1 = {\bar v}_{\rm a} \rho_{\rm a}+ {\bar v}_{\rm b} \rho_{\rm b}
\quad {\rm and}\quad s={\bar s}_{\rm a} \rho_{\rm a}+ {{\bar s}_{\rm b}} \rho_{\rm b}
\label{eqn:rhovv}\end{equation}
at equilibrium with $\rho_n$ and $s$ being homogeneous.  
Writing ${\bar v}_-$ and ${\bar s}_-$
for $({\bar v}_{\rm a}-{\bar v}_{\rm b})/2$ and $({\bar s}_{\rm a}-{\bar s}_{\rm b})/2$, respectively,
we obtain 
\begin{equation}
\delta \mu_-=  - {\bar s}_-^{({\rm ref})}\delta T+ {\bar v}_-^{({\rm ref})}\delta P+
\left.\frac{\partial \mu_- }{\partial c_{\rm a}}\right)_{T, P}\ \delta c_{\rm a}\ ,\label{eqn:kakkoP2}
\end{equation}
where the partial derivative is evaluated in the middle of a reservoir in the reference state.
Thus, the thermodynamic forces in Eq.~(\ref{eqn:Theta}) are {determined} by
$\delta T$, $\delta P$, and $\delta c_{\rm a}$, and determine $\delta \mu_+$ because
the {GD} relation gives
\begin{equation}
\delta P= \rho_n^{({\rm ref})} \delta \mu_n+s^{({\rm ref})} \delta T=
\rho^{({\rm ref})}\delta\mu_+ +\varphi^{({\rm ref})} \delta\mu_- +s^{({\rm ref})} \delta T
\ .\label{eqn:deltaP}\end{equation}

\subsection{Hydrodynamics\label{sec:hydro}}
{We} consider the Helmholtz free-energy of a mixture
as a functional of fields coarse-grained up to the local correlation length of the order-parameter fluctuations.
The length is denoted by $\xi$.  {A part of the functional
is} given by the volume integral of a function 
over the mixture region, $V_{\rm tot}$.
The function, denoted by $f_{\rm bulk}$, is assumed to depend on
$\rho_{\rm a}$, $\rho_{\rm b}$, their gradients, and $T$,
with the dependence on the gradients being via $|\nabla \rho_{\rm a}|^2$,
$|\nabla \rho_{\rm b}|^2$ and $(\nabla \rho_{\rm a})\cdot(\nabla\rho_{\rm b})$.   
The other part, representing the wall-component interaction, 
is assumed to be given by the area integral of a function over the interface, $\partial V_{\rm tot}$.  
This function is denoted by $f_{\rm surf}$ and depends on $\rho_{\rm a}$, $\rho_{\rm b}$, and $T$. 
Thus, the free-energy functional is given by
\begin{equation}
F[T, \rho_{\rm a}, \rho_{\rm b}]=\int_{V_{\rm tot}}d{\v r}\ 
f_{\rm bulk} \left(T, \rho_{\rm a}, \rho_{\rm b}, \nabla\rho_{\rm a}, \nabla\rho_{\rm b} \right)
+ \int_{\partial V_{\rm tot}}dA\ f_{\rm surf}\left(T, \rho_{\rm a}, \rho_{\rm b}\right)
\ ,\label{eqn:general}\end{equation}
where $T$ and $\rho_n$ depend on the position ${\v r}$. 
The volume element and the area element are represented by $d{\v r}$ 
and $dA$, respectively. \\

\noindent
We can formulate hydrodynamics on length scales larger than $\xi$ by using Eq.~(\ref{eqn:general}). 
If $T$ is homogeneous,
$\mu_n({\v r})$ is given by  the functional derivative of the first term on the RHS of Eq.~(\ref{eqn:general})
with respect to $\rho_n({\v r})$ in $V_{\rm tot}$.  Otherwise, it is given by
\begin{equation}
\mu_n=\frac{\partial f_{\rm bulk}}{\partial\rho_n}-T\nabla\cdot\left[ \frac{1}{T}\frac{\partial f_{\rm bulk}}{\partial\left(\nabla
\rho_n\right)}\right]\ .\label{eqn:chempot}\end{equation}
The reversible part of the pressure tensor, denoted by ${\mathsf \Pi}$, is given by
\begin{equation}
{\mathsf \Pi}=P {\mathsf 1}+ 
\left(\nabla \rho_n\right) \frac{\partial f_{\rm bulk}}{\partial\left(\nabla\rho_n\right)}
\ ,\label{eqn:Pidef}\end{equation} which is symmetric.  Here,
${\mathsf 1}$ is the identity tensor of order two. The scalar pressure $P$ is given by 
the negative of the grand-potential density,
\begin{equation}
P=  \mu_n\rho_n-f_{\rm bulk}=
\mu_+\rho+\mu_-\varphi-f_{\rm bulk}
\ .\label{eqn:scalarP}\end{equation}
Equations (\ref{eqn:chempot}) and (\ref{eqn:Pidef})  
are derived for a one-component fluid in Ref.~\cite{dvw}
and are applied in a straightforward way to a binary fluid mixture \cite{gonn}. 
The previous derivation is not applicable to
the free-energy density introduced in Section \ref{sec:free} 
because the coefficient of its square-gradient term, $M_-/2$ in $f_{\rm bulk}$ of Eq.~(\ref{eqn:efu}), 
{depends on $T$ slightly nonlinearly.  The linear dependence} is assumed in the previous derivation.
In {Appendix \ref{sec:stress}, we show a more general derivation, which is applicable to $f_{\rm bulk}$} of Eq.~(\ref{eqn:efu}).
{Strictly speaking, this derivation remains relevant to} our later calculation of 
thermoosmosis under the linear regime, because
the hydrodynamic equations including this nonlinearity must be derived before their linearization with respect to the temperature gradient.} 
Equations (\ref{eqn:chempot}) and (\ref{eqn:Pidef}) yield {an extended GD formula}
\begin{equation}
{\nabla\cdot {\mathsf\Pi}=\nabla\cdot {\mathsf\Pi}^{\rm T}
=\rho_n\nabla\mu_n +s\nabla T + \frac{\nabla T}{T}\cdot 
\frac{\partial f_{\rm bulk}}{\partial \left(\nabla \rho_n\right)} \nabla\rho_n\ ,}\label{eqn:nablaPi}
\end{equation}
which {is consistent with principles of linear nonequilibrium thermodynamics; 
Eq.~(\ref{eqn:nablaPi}) guarantees positive entropy production rate after combined with irreversible terms and
 the Onsager's reciprocity  for osmotic fluxes through the tube.  
The former is shown in Ref.~\cite{dvw},
whereas the latter is} {mentioned below Eq.~(\ref{eqn:calK}) in the next subsection}.
\\

\noindent
The velocity field, ${\v v}$, is defined in the frame fixed to the container.
The time-derivative of $\rho$ equals $-\nabla\cdot (\rho{\v v})$.  {In the stationary state}, we have
\begin{equation}
0=\nabla\cdot \left(\rho {\v v}\right) \ .\label{eqn:renzoku}
\end{equation}
Because of  the mass conservation of each component,
the time derivative of $\rho_n$ is equal to the negative of the divergence of its flux,
whose deviation from the convective part, $\rho_n {\v v}$, gives the diffusion flux, denoted by ${\v j}_n$.
It is defined so that ${\v j}_{\rm a}+{\v j}_{\rm b}$ vanishes.  In the stationary state, 
we have
\begin{equation}
0=\nabla\cdot \left( \varphi {\v v}+{\v j}\right) 
\ ,\label{eqn:diffusion}\end{equation}
where ${\v j}$ is defined as ${\v j}_{\rm a}-{\v j}_{\rm b}$. 
\\

\noindent
Assuming that  $\delta T$, $\delta P$, and $\delta c_{\rm a}$
are proportional to a dimensionless smallness parameter, $\varepsilon$,
we consider the dynamics in the tube at 
{the order of $\varepsilon$ in Section \ref{sec:tubefield}\@.}
The superscripts $^{(0)}$ and $^{(1)}$ are used to indicate the order of $\varepsilon$.
For example, we have
$\mu_\pm=\mu_\pm^{(0)}+\varepsilon \mu_\pm^{(1)}$ up to the order of $\varepsilon$,
$\mu_\pm^{(0)}=\mu_\pm^{({\rm ref})}$, and $T^{(0)}=T^{({\rm ref})}$.
In contrast, $\rho_n^{(0)}$ {becomes} inhomogeneous and different from $\rho_n^{({\rm ref})}$ in the presence of PA.
Because the fields are assumed to be coarse-grained up to $\xi$,
the mass densities at {the} equilibrium specified by $T^{(\rm ref)}$ and $\mu_n^{({\rm ref})}$
minimize the grand-potential functional,
\begin{equation} {F}[T^{({\rm ref})}, \rho_{\rm a}, \rho_{\rm b}]
- \int_{V_{\rm tot}}d{\v r} \ \left[\mu_+^{({\rm ref})} \rho({\v r})+\mu_-^{({\rm ref})} \varphi({\v r})\right]
\ .\label{eqn:relation}\end{equation}  
Thus, {$\rho_n^{(0)}$} is the solution of
Eq.~(\ref{eqn:chempot}) with $T$ and $\mu_n$ being replaced by $T^{({\rm ref})}$ and $\mu_n^{({\rm ref})}$,
respectively, together with the boundary conditions given by Eq.~(\ref{eqn:bc}).

\subsection{{Fields in the tube}\label{sec:tubefield}}
{In this subsection, we consider stationary and laminar flow in the tube at the order of $\varepsilon$. 
The mixture is assumed to remain in one-phase region throughout inside the container. 
As in the previous study \cite{pipe}, we assume
that the tube is so long and thin that effects of the} tube's {edges on the flow are negligible.
We regard $\delta \mu_n$ and $\delta T$ as 
equal to the differences in $\mu_n$ and $T$ between the} edges, respectively; 
{$\mu_n$ and $T$} are assumed to be homogeneous over the tube's cross-section at each edge.
\\

\noindent
Assuming the tube to extend along the $z$ axis with the same cross-section, we take
the Cartesian coordinates $(x,y,z)$ with the right reservoir lying on the positive $z$ side.
{A field with the superscript $^{(0)}$, such as $\rho_n^{(0)}$, is independent of $z$ in the tube.
We apply Eq.~(\ref{eqn:renzoku}) for a laminar flow to obtain} 
\begin{equation}
0=\nabla\cdot {\v v}^{(1)}=\partial_z v_z^{(1)}\ ,
\label{eqn:nablavtube}\end{equation}
without assuming the mixture to be incompressible.
Here, $\partial_z$ denotes the partial derivative with respect to $z$. 
In the tube up to the order of $\varepsilon$, the momentum conservation gives
\begin{equation}
2\nabla\cdot\left(\eta_{\rm s}{\mathsf E}
\right)=\nabla\cdot {\mathsf \Pi}\ ,
\label{eqn:stok}
\end{equation}
where $\eta_{\rm s}$ is the shear viscosity and ${\mathsf E}$ is the rate-of-strain tensor.
Owing to the critical enhancement, 
$\eta_{\rm s}$ depends on the local correlation length, $\xi$.
Writing $\eta_{\rm s}^{(0)}$
for $\eta_{\rm s}$ evaluated at $\varepsilon=0$, we rewrite the LHS of Eq.~(\ref{eqn:stok}) as
$\varepsilon$ multiplied by
\begin{equation}
2\nabla\cdot\left(\eta_{\rm s}^{(0)}{\mathsf E}^{(1)}\right)=
\nabla\cdot {\left\{ \eta_{\rm s}^{(0)} \left[\nabla {\v v}^{(1)}+\left(\nabla {\v v}^{(1)}\right)^{\rm T}\right]\right\}}
\ .\label{eqn:stok1L}\end{equation} 
With the aid {of} Eq.~(\ref{eqn:nablaPi}), we rewrite the RHS of Eq.~(\ref{eqn:stok}) as
$\varepsilon$ multiplied by
\begin{equation}
\nabla\cdot {\mathsf \Pi}^{(1)}= 
\rho^{(0)}_n\nabla \mu_n^{(1)} 
+s^{(0)}\nabla T^{(1)}
+\frac{\nabla T^{(1)}}{T^{(0)}}\cdot\frac{\partial f_{\rm bulk}}{\partial (\nabla \rho_n)}  \left(\nabla \rho_n^{(0)}\right)
\ ,\label{eqn:stok1R}\end{equation}
where the partial derivative of $f_{\rm bulk}$ is evaluated at $\varepsilon=0$.
In the absence of PA, $\rho_n^{(0)}$ and $\eta_{\rm s}^{(0)}$ are homogeneous, and thus 
Eq.~(\ref{eqn:stok}) becomes the usual Stokes equation,
$\eta_{\rm s}\Delta {\v v}=\nabla P$, owing to Eq.~(\ref{eqn:Pidef}).\\

\noindent
Writing $u$ ($e$) for the internal (total) energy per unit volume, we have $u=e-\rho \left|{\v v}\right|^2/2$,
\begin{equation}
s=-\left.\frac{\partial f_{\rm bulk}}{\partial T}\right)_{\rho_n, \nabla\rho_n}\ ,\  {\rm and}\quad 
u=f_{\rm bulk}+Ts=-T^2 \left.\frac{\partial}{\partial T}\frac{f_{\rm bulk}}{T}\right)_{\rho_n, \nabla\rho_n}
\label{eqn:sandu}\end{equation}
in {$V_{\rm tot}$.}
The heat flux, denoted by ${\v j}_q$,
is defined so that the {Eulerian} time-derivative of $e$ equals 
$-\nabla\cdot(e{\v v}+{\v v}\cdot {\mathsf \Pi}+{\v j}_q)$.  In a stationary state {in the tube}, we have 
\begin{equation}
0=\nabla\cdot\left( u^{(0)}{\v v}^{(1)}+ {\v v}^{(1)}\cdot {\mathsf \Pi}^{(0)}+ {\v j}_{\rm q}^{(1)} \right)\ ,\label{eqn:eneconv}
\end{equation}
where $u^{(0)}$ equals $e^{(0)}$.
We can define the transport coefficients, $\Lambda$, $\lambda$, and $\kappa$, so that 
\begin{equation}
{\v j}=-T\Lambda \nabla \frac{\mu_-}{T}+\kappa \nabla\frac{1}{T}
\quad{\rm and}\quad
{\v j}_{\rm q}=-\kappa \nabla \frac{\mu_-}{T}+\lambda 
 \nabla\frac{1}{T}\label{eqn:linear1}
\end{equation}
hold \cite{gromaz, engineer}.
The coefficients depend on $\xi$ {owing to the critical enhancement}.  
If evaluated using $\xi$ at $\varepsilon=0$,  
they are denoted by $\Lambda^{(0)}$, $\kappa^{(0)}$, and $\lambda^{(0)}$, respectively,
and are independent of $z$.
In Eq.~(\ref{eqn:linear1}) at the order of $\varepsilon$, 
we can use $T^{(0)}\Lambda^{(0)}$, $\kappa^{(0)}$, and $\lambda^{(0)}$ for
$T\Lambda$, $\kappa$, and $\lambda$, respectively. \\

\noindent
With the aid of Eqs.~(\ref{eqn:stok1L}) and (\ref{eqn:stok1R}), the $x$ and $y$ components of Eq.~(\ref{eqn:stok}) give
\begin{equation}
0=\rho^{(0)}{\bar\nabla} \mu_+^{(1)}+\varphi^{(0)} {\bar\nabla} {\mu_-^{(1)}}+s^{(0)}{\bar \nabla}T^{(1)}\ ,
\label{eqn:ppvpmxy}\end{equation} 
where ${\bar \nabla}$ represents the two-dimensional nabla defined on the $(x, y)$ plane.
Because $P^{(0)}$, $\rho_n^{(0)}$, and $u^{(0)}$ are independent of $z$, we obtain
\begin{equation}
\nabla\cdot{\v j}^{(1)}=0\quad{\rm and}\quad \nabla\cdot{\v j}_q^{(1)}=0\ .\label{eqn:jjq}
\end{equation}
The first entry comes from Eqs.~(\ref{eqn:diffusion}) and (\ref{eqn:nablavtube}), whereas
the second from Eqs.~(\ref{eqn:Pidef}), {(\ref{eqn:nablaPi})}, and (\ref{eqn:eneconv}).
The components of ${\v j}$ and ${\v j}_q$
normal to the tube's wall vanish {at the impermeable and adiabatic wall}.  
These conditions, the conditions at the tube's edges
mentioned in {the beginning of Section \ref{sec:tubefield}}, Eq.~(\ref{eqn:ppvpmxy}), and Eq.~(\ref{eqn:jjq})
are satisfied {if}
$\mu_n^{(1)}$ and $T^{(1)}$ are linear functions of $z$
and are independent of $x$ and $y$.  {Then}, ${\v j^{(1)}}$ and ${\v j}_q^{(1)}$ have
only $z$ components and are independent of $z$, considering Eq.~(\ref{eqn:linear1}) up to the order of $\varepsilon$.
With the aid of Eqs.~(\ref{eqn:stok1L}) and (\ref{eqn:stok1R}), the $z$ component of Eq.~(\ref{eqn:stok}) give
\begin{equation}
{\bar \nabla}\cdot \left({\eta}_{\rm s}^{(0)} {\bar \nabla} v_z^{(1)}\right)
=\rho^{(0)}_n\partial_z\mu_n^{(1)}
+s^{(0)}\partial_z T^{(1)}
\ ,\label{eqn:st3xy}\end{equation}
where $v_z^{(1)}$ is regarded as a scalar on a cross-section.   
The derivatives on the RHS above {are constants determined by
the thermodynamic forces in Eq.~(\ref{eqn:Theta})}.
Writing $L_{\rm tube}$ for the tube length, we obtain
\begin{equation}
\varepsilon \partial_z T^{(1)}=-\frac{T^{(0)2}}{L_{\rm tube}}\delta \left(\frac{1}{T}\right)
\ ,\quad 
\varepsilon \partial_z \mu_-^{(1)} =\frac{T^{(0)}}{L_{\rm tube}}\left[\delta \left(\frac{\mu_-}{T}\right)- 
\mu_-^{(0)} \delta \left(\frac{1}{T}\right)\right] 
\ ,\end{equation}
and
\begin{equation}
\varepsilon \partial_z \mu_+^{(1)}
=\frac{1}{\rho^{({\rm ref})}L_{\rm tube}}\left(\delta P-\varphi^{({\rm ref})}\delta\mu_--s^{({\rm ref})}\delta T\right)\ ,
\end{equation}
with the aid of Eq.~(\ref{eqn:deltaP}). 
Thus, we use Eqs.~(\ref{eqn:scalarP}) and (\ref{eqn:sandu}) to rewrite
Eq.~(\ref{eqn:st3xy}) as
\begin{eqnarray}&&
\varepsilon {\bar \nabla}\cdot \left({{\eta}_{\rm s}^{(0)}} {\bar \nabla} v_z^{(1)}\right)
=\frac{T^{(0)}}{L_{\rm tube}}\left\{
\frac{\rho^{(0)}}{\rho^{({\rm ref})}}\delta\left(\frac{P}{T}\right)+ \left( \varphi^{(0)}-\frac{\rho^{(0)}\varphi^{({\rm ref})}}{\rho^{({\rm ref})}} 
\right)\delta \left(\frac{\mu_-}{T}\right) \right.\nonumber\\&& \qquad+\left.
\left(\frac{\rho^{(0)}u^{({\rm ref})}}{\rho^{({\rm ref})}}-u^{(0)}-P^{(0)}\right)\delta\left(\frac{1}{T}\right)\right\}
\ .\label{eqn:st3xy+}\end{eqnarray}
This determines $v_z^{(1)}$, which is independent of $z$ because of Eq.~(\ref{eqn:nablavtube}), together with
{the boundary conditions}.
From Eq.~(\ref{eqn:linear1}), we obtain
\begin{equation}
\varepsilon j_z^{(1)}=\frac{1}{L_{\rm tube}} \left[T^{(0)}\Lambda^{(0)}\delta\left(\frac{-\mu_-}{T}\right)+\kappa^{(0)} \delta\left(\frac{1}{T}\right)\right]
\label{eqn:jzprofile}\end{equation}
and
\begin{equation}
\varepsilon j_{qz}^{(1)}=
\frac{1}{L_{\rm tube}} \left[\kappa^{(0)}\delta\left(\frac{-\mu_-}{T}\right)+\lambda^{(0)} \delta\left(\frac{1}{T}\right)\right]
\label{eqn:jqzprofile}\end{equation}
in the tube.  Up to the order of $\varepsilon$, $d{\cal M}_{n {\rm R}}/(dt)$ and $d{\cal U}_{\rm R}/(dt)$ 
are respectively given by 
the area integral of $\varepsilon(\rho_{n}^{(0)} v_z^{(1)} +j_{{n}z}^{(1)})$
and that of $\varepsilon(u^{(0)}+P^{(0)}) v_z^{(1)} +\varepsilon j_{q z}^{(1)}$ over a tube's cross-section, which is denoted by 
$S_{\rm tube}$.
Thus, we use the first equality of Eq.~(\ref{eqn:Theta}) to obtain
\begin{equation}
{\cal I}=\frac{\varepsilon}{\rho^{({\rm ref})}}  \int_{S_{\rm tube}}dA\ 
\rho^{(0)}v_z^{(1)}\ ,
\quad {\cal J}=\varepsilon  \int_{S_{\rm tube}}dA\ \left[\left(\varphi^{(0)}-\frac{\rho^{(0)}\varphi^{({\rm ref})}}{\rho^{({\rm ref})}}\right)
v_z^{(1)}+j_z^{(1)}\right]\ ,
\label{eqn:calIcalJ}\end{equation}
and
\begin{equation}
{\cal K}= \varepsilon  \int_{S_{\rm tube}}dA\  \left[\left(u^{(0)}+P^{(0)}-\frac{\rho^{(0)}u^{({\rm ref})}}{\rho^{({\rm ref})}}
\right) v_z^{(1)} +j_{q z}^{(1)}\right]
\ .\label{eqn:calK}\end{equation}
In Appendix \ref{sec:incep}, 
our formulation up to here is shown to be consistent with
Onsager's reciprocal relation, as it should be.
Because of Eq.~(\ref{eqn:scalarP}), $P^{(0)}$ can be inhomogeneous on a tube's cross-section 
in the presence of PA.  In its absence, because $P^{(0)}$ is homogeneously equal to $P^{({\rm ref})}$, 
the RHS of Eq.~(\ref{eqn:st3xy+}), and thus
$v_z^{(1)}$, vanish when $\delta P$ vanishes.
This is consistent with the results in Chapter XV-5 of Ref.~\cite{gromaz}.
\\

\noindent
{With the subscript $_{\rm c}$, we refer to the value at the critical point determined under 
the} pressure {$P^{({\rm ref})}$}.
The deviation of $\varphi$ from its value at the critical point,
$\varphi-\varphi_{{\rm c}}$, plays a role of the order parameter of  phase separation and is denoted by $\psi$.  
At equilibrium, 
correlated clusters of $\psi$ are randomly convected on length scales smaller than $\xi$.
On larger length scales, the convection is averaged out to enhance the transport coefficient 
for the interdiffusion {in a mixture at the critical composition, as mentioned in Section \ref{sec:intro}.
The critical enhancement
suppresses} the critical slowing down of the relaxation of the two-time correlation function of $\psi$. This function  
follows the diffusion equation.
According to the mode-coupling theory \cite{kawasaki},  the singular part of
the diffusion coefficient coincides with the self-diffusion coefficient of a rigid sphere
with the radius being equal to $\xi$.  This result is slightly modified by the dynamic renormalization-group
calculation for the model H, with the weak singularity of $\eta_{\rm s}$ taken into account \cite{halhohsig, ohta, sighalhoh, sengers, folkmos}.  
Because multiplying the diffusion coefficient by the osmotic susceptibility, 
denoted by $\chi$, gives the transport coefficient $\Lambda$, we have
\begin{equation}
\Lambda=\frac{{\chi} Rk_{\rm B} T_{\rm c}}{\xi \eta_{\rm sing}}
\ ,\label{eqn:Lambda0}\end{equation} 
where $R$ is a universal constant close to $1/(6\pi)$, $k_{\rm B}$ is the Boltzmann constant,
$T_{\rm c}$ is the critical temperature, and $\eta_{\rm sing}$ represents the singular part of $\eta_{\rm s}$.
The regular part of $\Lambda$ is usually negligible in the critical regime, judging from the data in Ref.~\cite{mirz}
for example.  The partial enthalpy per unit mass of the component $n$, denoted by ${\bar H}_n$, is given by
\begin{equation}
{\bar H}_n=\mu_n+T{\bar s}_n=-T^2\left.\frac{\partial}{\partial T} \frac{\mu_n}{T} \right)_{P,c_{\rm a}}
\label{eqn:barHn}\end{equation}
and ${\bar H}_-$ is defined as $({\bar H}_{\rm a}-{\bar H}_{\rm b})/2$.
{Neglecting the regular parts in the results of Ref.~\cite{mist}, we use}
\begin{equation}
\kappa=\Lambda T{\bar H}_-\quad {\rm and}\quad \lambda=\Lambda T \left({\bar H}_-\right)^2
\ .\label{eqn:kaplam}\end{equation} 
Here, as shown in Appendix \ref{sec:diss}, the second entry
contains more approximations, which 
affect only the formula for $L_{33}$ in the following, than the first entry. 
\\

\noindent
Equation (\ref{eqn:Lambda0}) holds at equilibrium with the critical composition.  In our problem, to evaluate $\Lambda^{(0)}$, 
we simply extend this result to a homogeneous off-critical composition
and use the extended result even when the composition is inhomogeneous.
{Hence, we  evaluate the RHS of Eq.~(\ref{eqn:Lambda0}) by using $T^{(0)}$, $\psi^{(0)}({\v r})$, and
the resulting local value of $\xi$, {to} obtain  $\Lambda^{(0)}$.}  This is the same procedure as used in Refs.~\cite{yabufuji, pipe}.
Likewise, we obtain $\kappa^{(0)}$ and $\lambda^{(0)}$ {in the dynamics} 
by replacing $\Lambda$, {$T$},  and ${\bar H}_-$ with $\Lambda^{(0)}$, {$T^{(0)}$}, 
and ${\bar H}_-^{(0)}$, respectively,
in Eq.~(\ref{eqn:kaplam}).  

\section{Calculation under some specifications \label{sec:proc}}
We specify the problem by making {the following} assumptions.
First, we assume 
$f_{\rm surf}$ to be independent of $\rho$. 
Thus, $\rho^{(0)}$ equals $\rho^{({\rm ref})}$ homogeneously. 
Second, we assume {$\psi^{({\rm ref})}=0$.}  
Third, we assume the tube to be a cylinder with the radius of $r_{\rm tube}$.  {In the tube},
a field depends only on the distance from the central axis, $r$, on a cross-section, and
we can write $\psi^{(0)}(r)$, $\eta_{\rm s}^{(0)}(r)$, $v_z^{(1)}(r)$, and $j_z^{(1)}(r)$, for example.
{The} LHS of Eq.~(\ref{eqn:st3xy+}) becomes
$\varepsilon r^{-1}\partial_r\left( r{\eta_{\rm s}^{(0)}} \partial_r v_z^{(1)} \right)$;
$v_z^{(1)}$ vanishes at $r=r_{\rm tube}$
owing to the no-slip condition and  $\partial_{r}v_{z}^{(1)}$ vanishes at $r=0$ 
owing to the axissymmetry and smoothness of $v_z^{(1)}$.
Thus, {in the tube}, we obtain
\begin{eqnarray}&&
\varepsilon v_{z}^{(1)}(r)=
\int_{r}^{r_{\rm tube}}dr_1\ \frac{1}{ r_1{\eta_{\rm s}^{(0)}(r_1)}}\int_{0}^{r_1}dr_2\ r_2
\nonumber\\&&\quad {\frac{T^{(0)}}{L_{\rm tube}}} 
\left[ \delta\left(\frac{-P}{T}\right)
+\psi^{(0)}(r_2)\delta\left(\frac{-\mu_-}{T}\right)
+\left(u^{(0)}(r_2)-u^{({\rm ref})}+P^{(0)}(r_2)\right)\delta\left(\frac{1}{T}\right)\right]
\label{eqn:vzprofile2}\end{eqnarray}
for general thermodynamic forces.
Substituting
Eqs.~(\ref{eqn:jzprofile}), (\ref{eqn:jqzprofile}), and (\ref{eqn:vzprofile2}) into Eqs.~(\ref{eqn:calIcalJ}) and (\ref{eqn:calK})
yields formulas for the components of $L$ in Eq.~(\ref{eqn:Theta}), {as shown in Section \ref{sec:formulae}}.

\subsection{Formulas for the Onsager coefficients \label{sec:formulae}}
{A} dimensionless radial distance ${\hat r}$ is defined as $r/r_{\rm tube}$.
{We define $T_*$ so that $\xi$ becomes $r_{\rm tube}$ for $\psi=0$ at $T=T_*$.}
A characteristic order parameter $\psi_*$ is defined so that
$\xi$ becomes  $r_{\rm tube}$  for $\psi=\psi_*$ 
at {$T=T_{\rm c}$}.  A dimensionless order-parameter at $\varepsilon=0$,
{${\hat \psi}^{(0)}({\hat r})$, is defined as} $\psi^{(0)}({\hat r}r_{\rm tube} )/\psi_*$.
A characteristic chemical potential, $\mu_*$, is defined as
\begin{equation}
\mu_*=\frac{k_{\rm B}{{T_*}}}{3u^*r_{\rm tube}^3 \psi_*}
\ ,\label{eqn:muast}\end{equation}
where $u^*$ is the scaled coupling constant at the Wilson-Fisher fixed point
and equals $2\pi^{2}/9$ at the one loop order. 
We define $\eta_*$ and $\Lambda_*$ as $\eta_{\rm sing}$ and $\Lambda$ at $\psi=0$ and {$T=T_*$}, respectively.
Dimensionless transport coefficients 
{${\hat \eta}({\hat r})$ and ${\hat \Lambda}({\hat r})$ are defined  
as $\eta_{\rm s}^{(0)}/\eta_*$ and 
$T^{(0)}\Lambda^{(0)}/(T_*\Lambda_*)$ evaluated at $r={\hat r}r_{\rm tube}$, respectively.}
The flow rate
of Hagen-Poiseulle flow of a fluid, with the viscosity being 
$\eta_*$, driven by the pressure gradient $\mu_*\psi_*/L_{\rm tube}$, is denoted by ${\cal I}_*$
and is given by
\begin{equation}
{\cal I}_*=\frac{ \pi r_{\rm tube}^4\mu_*\psi_* }{8\eta_* L_{\rm tube}}\ .
\end{equation}
\bigskip

\noindent
We define a functional $\Omega[g_1, g_2]$, where $g_1$ and $g_2$ {are} functions, as
\begin{equation}
\Omega[g_1, g_2]=16 \int _0^1
dq_1\ q_1 g_1(q_1) \int_{q_1}^1dq_2\ \frac{1}{q_2{\hat \eta}(q_2)}\int_0^{q_2}dq_3\ q_3g_2(q_3)
\ ,\end{equation}
which is found to be equal to $\Omega[g_2, g_1]$ by exchanging the order of integrals. 
We have
\begin{equation}
L_{11}=\frac{{\cal I}_*T^{(0)}}{\mu_*\psi_*}\Omega[1, 1]\ ,\quad
L_{12}=L_{21}=\frac{{\cal I}_*T^{(0)}}{\mu_*}\Omega[1, {\hat \psi}^{(0)}]\ ,
\label{eqn:1112}\end{equation}
and
\begin{equation}
L_{22}=\frac{{\cal I}_*\psi_*T^{(0)}}{\mu_*}\Omega[{\hat \psi}^{(0)}, {\hat \psi}^{(0)}]+
2\pi r_{\rm tube}^2 \frac{\Lambda_*T_*}{L_{\rm tube}}
\int_0^1d{\hat r}\ {\hat r}{\hat \Lambda}({\hat r})\ .\label{eqn:prel22}
\end{equation}
These three formulas are essentially the same as obtained in Ref.~\cite{pipe},
{where the diffusioosmotic conductance is calculated using $L_{12}$}.
Introducing 
\begin{equation}
{\hat Y}^{(0)}({\hat r})=\frac{1}{\mu_*\psi_*} \left[{u^{(0)}}-u^{({\rm ref})}+P^{(0)}\right]\quad {\rm and}\quad 
{{\hat H}_-^{(0)}({\hat r})=\frac{{\bar H}_-^{(0)}}{\mu_*} }
\ ,\label{eqn:hatY}\end{equation}
where {$u^{(0)}$, $P^{(0)}$, and ${\bar H}_-^{(0)}$ are evaluated at $r={\hat r}r_{\rm tube}$,} we obtain new formulas, 
\begin{equation}
{L}_{13}={L}_{31}={\cal I}_*T^{(0)}\Omega[1, {\hat Y}^{(0)}]
\ , \label{eqn:1331}\end{equation}
\begin{equation}
L_{23}=L_{32}={\cal I}_*\psi_*T^{(0)} \Omega[{\hat \psi}^{(0)}, {\hat Y}^{(0)}] +
2\pi r_{\rm tube}^2  \frac{\mu_*\Lambda_*T_*}{L_{\rm tube}}
\int_0^1d{\hat r}\ {\hat r}{\hat \Lambda}({\hat r}){\hat H}_-^{(0)}({\hat r})\ ,
\label{eqn:prel23}\end{equation}
and 
\begin{equation}
L_{33}={\cal I}_*\mu_*\psi_*T^{(0)} \Omega[{\hat Y}^{(0)}, {\hat Y}^{(0)}] 
+2\pi r_{\rm tube}^2 \frac{\mu_*^2\Lambda_*T_*}{L_{\rm tube}}
\int_0^1d{\hat r}\ {\hat r}{\hat \Lambda}({\hat r})\left[{\hat H}_-^{(0)}({\hat r})\right]^2 
\ .\label{eqn:prel33}\end{equation}
The second term on the RHS {of Eq.~(\ref{eqn:prel33})} 
is affected by the approximations mentioned below Eq.~(\ref{eqn:kaplam}).

\subsection{Formulas for thermoosmosis\label{sec:thermoosmo}}
The thermal force density occurs under a temperature gradient and causes thermoosmosis.
We write  $\sigma_z^{({\rm th})}$ for this density on a tube's cross-section in the linear regime. 
It is given by the negative of the RHS of Eq.~(\ref{eqn:st3xy+}), {or the $z$ component of 
$-\varepsilon \nabla\cdot\Pi^{(1)}$}, under $\delta T\ne 0$ and $\delta P=\delta c_{\rm a}=0$.
We have
\begin{equation}
\sigma_z^{({\rm th})}{(r)}=-\frac{\delta T}{T^{(0)} L_{\rm tube}}  
\left(u^{(0)}(r)+{P^{(0)}(r)-u^{({\rm ref})}}-P^{({\rm ref})}-\bar{H}_{-}^{({\rm ref})}\psi^{(0)}\right)\ .
\label{eqn:thermal-force-density}\end{equation}
The factor ${\bar H}_-^{({\rm ref})}$ above
comes from $\delta (\mu_{-}/T)$ via Eqs.~(\ref{eqn:kakkoP2}) and (\ref{eqn:barHn}).
The first four terms in the parentheses above 
can be interpreted as the excess enthalpy density in {DS's} formula {for a one-component fluid}. 
In the presence of PA, ${\mathsf \Pi}^{(0)}_{zz}=P^{(0)}$ is equal to neither
${\mathsf \Pi}^{(0)}_{xx}$ nor  ${\mathsf \Pi}^{(0)}_{yy}$.
The occurrence of ${\mathsf \Pi}^{(0)}_{zz}=P^{(0)}$ in Eq.~(\ref{eqn:thermal-force-density}) is 
consistent with many author's claim that the $zz$ component
should be involved in the thermal force density near the wall \cite{anders, ganti}.\\

\noindent
Here, we compare our derivation of the 
thermal force density with the corresponding part in Ref.~\cite{ganti}, 
which is mentioned in Section \ref{sec:intro}.  
Because the sum of the last three terms of Eq.~(\ref{eqn:thermal-force-density}) equals
{$-\rho_{\rm a}{\bar H}_{\rm a}^{({\rm ref})}-\rho_{\rm b}{\bar H}_{\rm b}^{({\rm ref})}$}, 
thus the negative of Eq.~(\ref{eqn:thermal-force-density}), {\it i.e.\/}, the formula for 
$-\sigma^{({\rm th})}_z$, formally coincides with the RHS of
{Eq.~(5)} {of Ref.~\cite{ganti}, where the RHS is treated as the negative of the thermal force density.
However, its LHS, {$\partial_z {\mathsf \Pi}_{zz}$ in our notation},  
is not equal to {$-\sigma^{({\rm th})}_z$} in general, 
since $\partial_x {\mathsf \Pi}_{xz}+\partial_y {\mathsf \Pi}_{yz}$ does not vanish 
in the presence of PA, up to the order of $\varepsilon$.
In Ref.~\cite{ganti}, this sum {$\partial_x {\mathsf \Pi}_{xz}+\partial_y {\mathsf \Pi}_{yz}$} 
is also missing in {the LHS of} Eq.~(2), which the authors employ
as an extended GD relation in deriving their Eq.~(5). 
In the present study, we use our Eq.~(\ref{eqn:nablaPi}), whose LHS includes 
$\partial_x {\mathsf \Pi}_{xz}+\partial_y {\mathsf \Pi}_{yz}$,
 as an extended GD relation to  derive our formula for the thermal force density 
$\sigma^{({\rm th})}_z$ of Eq.~(\ref{eqn:thermal-force-density}), 
consistently with principles of linear nonequilibrium thermodynamics. 
\\

\noindent
{The superscript $^{({\rm th})}$ is used to indicate a result} {in the linear regime} for thermoosmosis.
Replacing the integrand  in Eq.~(\ref{eqn:vzprofile2}) by $\sigma_z^{({\rm th})}(r_2)$ gives
$v_z^{({\rm th})}(r)$.
Integrating this result multiplied by $\rho^{({\rm ref})}$
over the tube's cross-section gives the total mass flow rate, for which we write $d{\cal M}_{\rm R}^{({\rm th})}/(dt)$.
Using {the free-energy functional introduced in Section \ref{sec:free}, we
rewrite Eq.~(\ref{eqn:thermal-force-density}) and give an explicit expression of $d{\cal M}_{\rm R}^{({\rm th})}/(dt)$
in Section \ref{sec:eval}}\@.
In the absence of PA, because Eq.~(\ref{eqn:thermal-force-density}) vanishes,
$v_z^{({\rm th})}$ {and $d{\cal M}_{\rm R}^{({\rm th})}/(dt)$} vanish.
Thus, in our formulation,
{thermoosmosis of a mixture occurs only in the presence of PA}.\\

\noindent
{Setting $\delta T\ne 0$ and $\delta P=\delta c_{\rm a}=0$}, we use 
Eqs.~(\ref{eqn:Theta}), (\ref{eqn:kakkoP2}), and (\ref{eqn:barHn}) to obtain 
the mass flow rate of the component $n$ {in thermoosmosis} as
\begin{equation}
\frac{d}{dt} {\cal M}_{n{\rm R}}^{({\rm th})}=
\frac{\delta T}{(T^{(0)})^2}\left[\rho_{n}^{({\rm ref})}
\left(P^{({{\rm ref}})}L_{11}+{\bar H}_-^{({\rm ref})}L_{12}-L_{13}\right)
 \pm \frac{1}{2} \left(P^{({{\rm ref}})}L_{21}+{\bar H}_-^{({\rm ref})}L_{22}-L_{23}\right)\right]
\ ,\label{eqn:massflow}\end{equation}
where the upper (lower) sign is taken for $n=$a (b) in the double sign. 
{The sum of Eq.~(\ref{eqn:massflow}) over $n=$a and b 
gives $d{\cal M}_{\rm R}^{({\rm th})}/(dt)$. Rewriting the resultant sum with the aid of
the formulas for $L_{11}$, $L_{12}$, and $L_{13}$ shown in Section \ref{sec:formulae},
we obtain the same expression of $d{\cal M}_{\rm R}^{({\rm th})}/(dt)$ as derived in the way mentioned in the preceding paragraph}.

\subsection{Free-energy functional in the renormalized local functional theory\label{sec:free}}
The reduced temperature $\tau$ is defined as $(T-T_{\rm c})/T_{\rm c}$,
and its characteristic magnitude $\tau_*$ is defined as $|T_*-T_{\rm c}|/T_{\rm c}$.
The scaled {reduced-temperature} ${\hat \tau}$ is defined as $\tau/\tau_*$. 
In {the one-phase region} we consider, $\tau$ is positive near a upper consolute (UC) point 
and is negative near a lower consolute (LC) point \cite{kaji, sciam, tsori}.
Using the conventional notation, we write $\alpha, \beta,\gamma,\nu,$ and $\eta$ for the critical exponents 
for a mixture.  We adopt $\nu= 0.630$ and $\eta= 0.0364$ \cite{peli};
the (hyper)scaling relations give $2\beta+\gamma=3\nu=2-\alpha$ and
$\gamma=\nu(2-\eta)$.   
{In} an equilibrium mixture with $\psi=0$, we have $\xi=\xi_0{\left|\tau\right|}^{-\nu}$
in the critical regime, where $\xi_0$ is a material {constant.} \\

\noindent
Neglecting coupling between $\rho$ and $\varphi$ 
in $f_{\rm bulk}$, we assume
\begin{equation}
f_{\rm bulk}= -{\frac{{C}T\tau^2}{2}}+{u}_{\rm c}-{s}_{\rm c}T+f_-(\psi)
+\frac{M_-}{2} \left|\nabla\psi\right|^2+f_+(\rho)\ .\label{eqn:efu}\end{equation}
Although the variable $\tau$ is dropped for conciseness,
$f_+$ is a regular function of $\rho$ and $\tau$ and $f_-$ is a function of $\psi$ and $\tau$.
The constants $u_{\rm c}$ and $s_{\rm c}$ represent the values of $u$ and $s$ at the critical point, respectively.  
{The coefficient $M_-$ is described later.}
In the critical regime, the singular contribution to 
$C$ \cite{onukibook, halhoma} becomes 
equal to $2k_{\rm B}\xi_0^{-3}{|\tau|^{-\alpha}}$
multiplied by a universal number, as mentioned at footnote 51 of Ref.~\cite{rlft}.\\

\noindent 
The reference state considered here is obtained by changing  
$T$ from the critical point with $P=P^{({\rm ref})}$ and $\psi=0$ being fixed.
{The chemical potentials, $\mu_{\rm a}^{({\rm ref})}$ and $\mu_{\rm b}^{({\rm ref})}$,
are tuned so that this change is realized.}
Thus, the $\varphi$ dependent part for the bulk of a mixture in Eq.~(\ref{eqn:relation})
can be obtained by coarse-graining the $\psi^4$ model up to $\xi$ 
under no external field \cite{rlft}. 
The bare model,  given by  Eq.~(\ref{eqn:bare}), is defined at a microscopic scale and 
identifies the fluctuations with spacial resolution much smaller than $\xi$.  
{We can regard the coarse-grained average profile as 
maximizing the probability density functional coarse-grained up to} $\xi$, 
{assuming that thermal fluctuations are not significant after coarse-graining  anymore} \cite{rlft}.
This is consistent with the statement given at Eq.~(\ref{eqn:relation}).
\\

\noindent
We assume no coupling between $\rho$ and $\varphi$ in
Eq.~(\ref{eqn:efu}) because $\rho$ can be regarded as a constant approximately. 
The coefficient $C$ involves the fluctuations of the internal-energy density.  
{Some details on these points}
 are mentioned in Appendix \ref{sec:fbulk}. 
 As shown in Eq.~(\ref{eqn:bare}), we can define $A_0$ so that the bare $\psi^4$ model has 
a term $A_0\tau \psi^2/2$, 
which is positive in the one-phase region. 
Thus, $A_0$ is positive near a UC point, and is 
negative near a LC point \cite{onukibook}.  
The sign is maintained in the coarse-graining procedure.
{We use the coarse-grained result} given by the renormalized local functional theory (RLFT) \cite{rlft}. 
\\

\noindent
In the RLFT, 
$\omega$ is defined as $(\xi_0/\xi)^{1/\nu}$ and $M_-$ is given by
$k_{\rm B}TC_1 \omega^{-\eta\nu}$ with $C_1(>0)$ being a material constant.
The self-consistent condition,
$\omega={\left|\tau\right|}+C_{2}\omega^{1-2\beta}\psi^{2}$, determines how $\xi$ depends on $\tau$ and $\psi$,
where the constant $C_2$ equals $3u^{*}C_{1}\xi_{0}$.  This condition gives
\begin{equation}
\tau_*=\left(\frac{\xi_{0}}{r_{\rm tube}}\right)^{1/\nu}\quad {\rm and}\quad 
\psi_*=\frac{\tau_*^\beta}{\sqrt{C_2}}
\ .\label{eqn:nodim}\end{equation}
Defining a dimensionless function ${\hat f}$ as
 \begin{equation}
{\hat f}({\hat\psi})={\frac{1}{2}
{\hat\omega}^{\gamma-1}
\left|{\hat \tau}\right|{\hat\psi}^2+\frac{1}{12}{\hat\omega}^{\gamma-2\beta}
{\hat\psi}^4}
\ ,\label{eqn:prefmm0}\end{equation}
where ${\hat\psi}\equiv \psi/\psi_*$, ${\hat \tau}\equiv \tau/\tau_*$, 
and ${\hat\omega}\equiv \omega/\tau_*$ are used,
we have
\begin{equation}
{f_-(\psi)+\frac{M_-}{2} \left|\nabla \psi\right|^2
=\mu_-^{({\rm ref})}\varphi + \frac{\mu_*\psi_*T}{T_*} {\hat f}({\hat \psi})+\frac{\mu_*\psi_*T }{2T_*}{\hat \omega}^{-\eta\nu}
\left| r_{\rm tube} \nabla {\hat\psi}\right|^2} 
\ .\label{eqn:prefmm} \end{equation}
{As already explained}, $\mu_-^{({\rm ref})}$ is determined so that the reference state at $T$ is 
realized, being dependent on $T$.   
The sum of the second and third terms on the RHS above is $k_{\rm B}T$ multiplied by
 the coarse-grained result of the $\psi^4$ model under no external field, coming from the RLFT.
The first term on the RHS of Eq.~(\ref{eqn:prefmm0}) originates  
{from $A_0\tau \psi^2/2$}. 
The self-consistent condition is rewritten as
\begin{equation}
{\hat \omega}={\left|{\hat \tau}\right|}+{\hat \omega}^{1-2\beta}{\hat \psi}^{2}
\ ,\label{eqn:omega2}\end{equation} 
{which means that ${\hat \omega}$ is a function of ${\hat \psi}$ and ${\hat \tau}$. 
It is even with respect to ${\hat \psi}$, 
and hence ${\hat f}({\hat \psi})$ is an even function.
The function} ${\hat f}({\hat\psi})$ also depends on ${\hat \tau}$, but 
the variable ${\hat \tau}$ is dropped for conciseness. 
{The osmotic susceptibility $\chi$ is given by the inverse of
the second partial derivative of $f_-$ with respect to $\psi$}, $1/f_-''(\psi)$;
the prime indicates the differentiation with respect the variable given explicitly.}
The partial derivative $\partial {\hat f}/(\partial {\hat \tau})$, appearing
in the later calculation, equals
\begin{equation}
{\frac{\pm 1}{2} {\hat\omega}^{\gamma-1}{\hat\psi}^2+\frac{\partial {\hat\omega}}{\partial {\hat\tau}}}
\left(\frac{\gamma-1}{2}{\hat \omega}^{\gamma-2}  \left|{\hat \tau}\right|{\hat\psi}^2
+\frac{\gamma-2\beta}{12} {\hat\omega}^{\gamma-2\beta-1} {\hat\psi}^4 \right)
\ ,\label{eqn:hatf}\end{equation}
where Eq.~(\ref{eqn:omega2}) gives
\begin{equation}
 \frac{\partial {\hat\omega}}{\partial {\hat\tau}}=\frac{\pm1}{1+(2\beta-1){\hat \omega}^{-2\beta}{\hat \psi}^2}
\ .\label{eqn:omegatau}\end{equation}
The same sign as $\tau$ is taken
in each double sign of these equations.
The first term of Eq.~(\ref{eqn:hatf}) originates from the coarse-grained result of $A_0\psi^2/2$.
{Equation (\ref{eqn:omega2})
gives $|{\hat\psi}|^{1/\beta}<{\hat \omega}$ and $|{\hat \tau}|<{\hat\omega}$.
For ${\hat\psi}\ne 0$, {the sign of Eq.~(\ref{eqn:hatf}), or that of $\partial {\hat f}/(\partial {\hat \tau})$, 
coincides with that of $\tau$, considering $\beta=0.326$ and $\gamma=1.24$.}
If $|{\hat \tau}|$ is much smaller than ${\hat\omega}$, Eq.~(\ref{eqn:omega2}) gives $|{\hat\psi}|^{1/\beta}\approx {\hat\omega}$
and thus Eq.~(\ref{eqn:omegatau}) is approximately equal to $\pm 1/(2\beta)$.
Then, in Eq.~(\ref{eqn:hatf}), the first term is} found {to be dominant over the rest.
If $|{\hat \tau}|$ is close to ${\hat\omega}$}, 
Eq.~(\ref{eqn:omegatau}) is found to be close to $\pm 1$ with the aid of Eq.~(\ref{eqn:omega2}).
Then, {the first term remains dominant in Eq.~(\ref{eqn:hatf}),
accounting for approximately $80$\% of the total owing to}
{the numerator $\gamma-1=0.24$} in the parentheses.
\\

\noindent
We assume $f_{\rm surf}$ to be a
linear function of $\varphi$, or $\psi$, as usual 
in studying the PA \cite{diehl97, rlft}. 
The surface field $h$ is defined as the negative of the coefficient of $\psi$.
This assumption and this definition are involved 
in calculating the equilibrium profile, which is used in Section \ref{sec:num}.
{The calculation procedure is mentioned 
below Eq.~(\ref{eqn:relation})} and is the same as that of Ref.~\cite{rlft}.
Applying Eqs.~(\ref{eqn:efu}) and (\ref{eqn:prefmm}), 
we find that ${\hat \psi}^{(0)}({\hat r})$ is the solution of 
\begin{equation}
0={\hat f}'({\hat\psi})-{\frac{1}{2}\frac{\partial {\hat \omega}^{-\eta\nu}}{\partial{\hat\psi}}}
\left(\partial_{\hat r}{\hat\psi}\right)^2-{\hat \omega}^{-\eta\nu}\left(\partial_{\hat r}^2+
\frac{1}{{\hat r}}\partial_{\hat r}\right){\hat\psi}\quad {\rm for}\ {\hat r}<1\ ,\label{eqn:stationary}
\end{equation} 
together with the boundary condition at the wall,
$({\hat\psi}^{0)})'(1)={\hat h} {\hat \omega}^{\eta\nu}$.
Here, ${\hat \omega}$
is regarded as a function of ${\hat \psi}$ and ${\hat \tau}$ via Eq.~(\ref{eqn:omega2})
and {a scaled surface field ${\hat h}$ is defined as ${h T_* /\left(T\mu_*r_{\rm tube}\right)}$.}
These equations are shown in Appendix D of Ref.~\cite{pipe}; {${\hat \psi}^{(0)}$ is totally determined by $|{\hat \tau}|$
and ${\hat h}$.}
{If we change the sign of ${\hat h}$,  the sign of ${\hat \psi}^{(0)}$ changes with the magnitude remaining the same.}
{Notably, $\left|h\right|$ represents the strength of the PA and vanishes in its absence}.

\subsection{Formulas incorporating the RLFT \label{sec:eval}}
We apply Eq.~(\ref{eqn:efu}) and the results of the RLFT to 
rewrite the RHS of Eq.~(\ref{eqn:thermal-force-density}).
Owing to Eqs.~(\ref{eqn:sandu}) and (\ref{eqn:prefmm}), {we have}
\begin{equation}
u^{(0)}-u^{({\rm ref})}=-{\frac{\left(T^{(0)}\right)^2\mu_*\psi_*}{T_{\rm c}T_*}}\frac{\partial}{\partial\tau}
\left(\hat{f}+\frac{\hat{\omega}^{-\eta\nu}}{2}\left|\partial_{\hat{r}}{{\hat\psi}^{(0)}}\right|^{2}\right)
+\bar{H}_{-}^{({\rm ref})}{\psi^{(0)}}\ ,\label{eqn:energy-density}
\end{equation}
where the partial derivative with respect to $\tau$ is done with ${\hat \psi}$ fixed and is evaluated at $\varepsilon=0$.
{The first term on the RHS represents the difference in the internal energy density
involved in the coarse-grained result of the $\psi^4$ model.}
In deriving the second term, we drop one term, which is proportional to the thermal expansion coefficient.
This term gives negligibly small
contribution to our later numerical results, as described in Appendix \ref{sec:detail}\@.
Owing to Eqs.~(\ref{eqn:scalarP}), (\ref{eqn:efu}), and (\ref{eqn:prefmm}), we have
\begin{equation}
P^{(0)}-P^{({\rm ref})}=
-\frac{\mu_{*}\psi_{*}{T^{(0)}}}{T_{*}}{\left(\hat{f}+\frac{\hat{\omega}^{-\eta\nu}}{2}\left|\partial_{\hat{r}}\hat{\psi}\right|^{2}
\right)}
\ .\label{eqn:scalar-pressure}\end{equation}
{We} define {a scaled} thermal force density, {${\hat\sigma}_z^{({\rm th})}$, so that
Eq.~(\ref{eqn:thermal-force-density}) is rewritten} as
\begin{equation}
\sigma_z^{({\rm th})}({\hat r}r_{\rm tube})= \frac{\mu_{*}\psi_{*}\delta T}{{\tau_{*}}{T_{*}}L_{\rm tube}}{\hat\sigma}_z^{({\rm th})}({\hat r})
\label{eqn:normalized-thermal-force-density2}
\ ,\end{equation}
and have
\begin{equation}
{\hat\sigma}_z^{({\rm th})}=
{\tau_*}\left({\hat f}+\frac{1}{2{\hat\omega}^{\eta\nu}}\left|\partial_{\hat r}{\hat \psi}\right|^2\right)+
{\frac{T^{(0)}}{T_{\rm c}}}\left(\frac{\partial {\hat f}}{\partial{\hat \tau}}-\frac{\eta \nu}{2{\hat \omega}^{\eta\nu+1}}
{\frac{\partial{\hat\omega}}{\partial{\hat \tau}}}\left|\partial_{\hat r} {\hat\psi}\right|^2\right)
\ ,\label{eqn:thermal-force-density0}
\end{equation}
which is evaluated at $\varepsilon=0$.
The first term on the RHS of Eq.~(\ref{eqn:thermal-force-density0})
comes from Eq.~(\ref{eqn:scalar-pressure}), whereas the second term comes from the first term
on the RHS of Eq.~(\ref{eqn:energy-density}).
The last term in the  parentheses of Eq.~(\ref{eqn:thermal-force-density}) cancels out the last term on the RHS of
Eq.~(\ref{eqn:energy-density}).  Thanks to this cancellation, the thermal force density 
does not involve {${\bar H}_-^{({\rm ref})}$}. 
{Except for the factors $\tau_*$ and $T^{(0)}/T_{\rm c}$}, 
Eq.~(\ref{eqn:thermal-force-density0}) is determined by the scaled {reduced-temperature} ${\hat \tau}$
and the magnitude of the scaled surface field $|{\hat h}|$ in the framework of the RLFT\@.
{The magnitude of the sum in the second parentheses, in particular, 
is determined by $|{\hat \tau}|$ and $|{\hat h}|$ owing to Eqs.~(\ref{eqn:hatf}) and (\ref{eqn:omegatau})}. \\

\noindent
As mentioned in Section \ref{sec:thermoosmo},
{$\varepsilon v_z^{(1)}$ in thermoosmosis is given by}
\begin{equation}
v_z^{({\rm th})}({\hat r}r_{\rm tube})=
\frac{8{\cal I}_*\delta T}{T_*\tau_*\pi r_{\rm tube}^2}
\int_{\hat r}^1d{\hat r}_1\ \frac{1}{{\hat r}_1{\hat\eta}({\hat r}_1)}\int_0^{{\hat r}_1}d{\hat r}_2
\ {\hat r}_2{\hat \sigma}_z^{({\rm th})}\label{eqn:nodimv0}
\end{equation}
in the tube, whereas the total mass flow rate in thermoosmosis is given by
\begin{equation}
\frac{d{\cal M}_{\rm R}^{({\rm th})}}{dt}=\frac{\rho^{({\rm ref})}{\cal I}_*\delta T}{T_*\tau_*}
\Omega\left[1, {\hat \sigma}^{({\rm th})}_z\right]
\ ,\label{eqn:PL1}\end{equation}
which is proportional to $\delta T$.   The constant of proportionality 
represents the thermoosmotic conductance.
We define the {dimensionless} thermoosmotic conductance, denoted by ${\hat G}^{({\rm th})}$, as the quotient of
 the constant divided by $\rho^{({\rm ref})}{\cal I}_*/(T_*\tau_*)$, and have
\begin{equation}
{\hat G}^{({\rm th})}= \Omega[1, {\hat \sigma}_z^{({\rm th})}]\ .\label{eqn:conduct}
\end{equation}
{If we change the sign of ${\hat h}$, 
${\hat\sigma}_z^{({\rm th})}({\hat r})$ remains the same and
is independent of which component is adsorbed onto the tube's wall. 
However, it is not the case with {${v}_z^{({\rm th})}$ of Eq.~(\ref{eqn:nodimv0})},  
$d{\cal M}_{\rm R}^{({\rm th})}/(dt)$ of Eq.~(\ref{eqn:PL1}),
{and} ${\hat G}^{({\rm th})}$ of Eq.~(\ref{eqn:conduct})
 because ${\hat \eta}$ 
is not always an even function of ${\hat \psi}$.\\

\noindent
Some of the formulas of the Onsager coefficients are simplified using Eq.~(\ref{eqn:efu}) and the results of the RLFT.
We have ${\hat f}''(0)={\left|{\hat\tau}\right|}^{\gamma}$ and
\begin{equation}
f_{-}''(0)=\frac{k_{\rm B}T C_2{\left|\tau\right|}^\gamma}{3u^*\xi_0^{3}}\ .
\label{eqn:fprpr}\end{equation}
Because of Eq.~(\ref{eqn:efu}), we can replace {$\chi$} in Eq.~(\ref{eqn:Lambda0}) with $1/f_-''(\psi)$.
By definition, we have
\begin{equation}
\Lambda_*=\frac{3u^*R{T_{\rm c}}\xi_0^2\tau_*^{\nu-\gamma}}{C_2{T_*}\eta_*} 
\ .\label{eqn:LamcalI}\end{equation}
{We write} $z_\psi$ for the dynamic critical exponent for the
order-parameter fluctuations and use $z_\psi=3.067$ \cite{bergmold, bergmold2}. 
{With the aid of the} result of the 
dynamic renormalization-group calculation \cite{sighalhoh, sengers, folkmos}, we have
\begin{equation}
{{\hat \Lambda}=}{\hat \omega}^{\nu(z_\psi-2)}\left[ {\hat f}''({\hat \psi})\right]^{-1}
\label{eqn:hatLam}\end{equation}
evaluated at $\varepsilon=0$, 
as described in Appendix E of Ref.~\cite{pipe}. 
Using Eq.~(\ref{eqn:Lambda0}) with $R=1/(6\pi)$ and
Eq.~(\ref{eqn:hatLam}), 
we can rewrite the second {term on the RHS of
Eq.~(\ref{eqn:prel22})} as 
\begin{equation}
\frac{{\cal I}_*\psi_*T^{(0)}}{\mu_*}
\frac{16\pi T_{\rm c} }{9T^{(0)}}
\int_0^1d{\hat r}\ {\hat r}{\hat \Lambda}({\hat r})
\ ,\label{eqn:prel22b}
\end{equation}
which is essentially the same as obtained in Ref.~\cite{pipe}.
Likewise, in the second terms on the RHS of the new formulas  (\ref{eqn:prel23}) and (\ref{eqn:prel33}),
the coefficients multiplied by the integrals are respectively rewritten as
\begin{equation}
{\cal I}_*\psi_*T^{(0)}
\frac{16\pi T_{\rm c}}{9T^{(0)}}
\quad{\rm and}\quad 
{\cal I}_*\mu_*\psi_*T^{(0)}
\frac{16\pi T_{\rm c} }{9T^{(0)}}
\ .\label{eqn:prel33b}\end{equation}
{The integrals can be calculated if ${\hat H}_-^{(0)}$ is known.
To calculate ${\hat Y}^{(0)}$ contained in the first terms on
the RHS's of Eqs.~(\ref{eqn:1331})--(\ref{eqn:prel33}), we can use Eq.~(\ref{eqn:energy-density}),
which involves ${\bar H}_-^{({\rm ref})}$}.  {Thus, it is necessary to know
how ${\bar H}_-$ depends on $\varphi$ to calculate these integrals and terms.
We can calculate the dependence in
such a theoretical framework as used in Refs.~\cite{edison,vanthof,olaya}.}

\section{Numerical results of thermoosmosis\label{sec:num}}
In this section, we study thermoosmosis {numerically with the aid of the formulas} in Section \ref{sec:eval} 
and the software Mathematica (Wolfram Research), supposing
a mixture of 2,6-lutidine and water (LW) near the LC point
and a mixture of nitroethane and 3-methylpentane
(NEMP) near the UC point.  In each mixture, the former (latter) component is 
taken to be the component a (b).
The tube radius $r_{\rm tube}$ is set to $0.1\ \mu$m.  
The parameter values we use are listed in Table \ref{tab:para} and are the same as used in Ref.~\cite{pipe}.
The values of $\xi_0$ 
are taken from the experimental {data of Refs.~\cite{mirz, iwan}.}
The first entry of Eq.~(\ref{eqn:nodim}) gives the values of $\tau_*$,
{which appears in Eq.~(\ref{eqn:thermal-force-density0}).}
In Appendix C of Ref.~\cite{pipe}, we estimate $C_2$
from the data of Refs.~\cite{to, wims}.  The second entry of Eq.~(\ref{eqn:nodim}) gives the value of $\psi_*$,
and then Eq.~(\ref{eqn:muast}) gives that of $\mu_*$.
In Appendix E of Ref.~\cite{pipe}, 
we obtain the viscosity as a function of the reduced temperature 
and the order parameter \cite{bhatt,tsai} and find the value of $\eta_*$
from the data of {Refs.~\cite{gratt,stein,iwan,leis}}.
These values give
${\cal I}_*L_{\rm tube}=1.04\times 10^{-2}\ \mu$m$^4/$s ($4.84\times 10^{-2}\ \mu$m$^4/$s)
for a mixture of LW (NEMP).  
{The} magnitude of the surface field $h$
should be smaller than approximately $10\ $cm$^3/$s$^2$ when
2,6-lutidine adsorbs onto the solid surface, according to the discussion in Section 6 of Ref.~\cite{yabufuji}.
We mainly use ${\hat h}=73.0$ ($66.6$) for a mixture of LW (NEMP), 
which amounts to {$h\approx 0.1\ $cm$^3$/s$^2$}.
\\

\begin{table}[t]
\begin{center}
\centering
\begin{tabular}{l | ccccccc}
\toprule
{mixture}\hspace{2mm} 
&\ ${T_{\rm c}}\ [{\rm K}]$\ & $\ \xi_0\ [{\rm nm}]\ $&  $\tau_*\times 10^5\ \ $ & $C_2 \ [{\rm cm}^6/{\rm g}^2]\ $&
$\psi_*\ [{\rm g}/{\rm cm}^3]\ $ & $\mu_*\ [{\rm cm}^2/{\rm s}^2]\ $&  
$\eta_*\ [{\rm mPa}{\cdot{\rm s}}]$ \\
\hline
LW  & 307  & 0.198& $5.12$ & $0.714$ & $0.0470$ & $137$ & $2.44$ \\
NEMP & 300 & 0.230 & $6.49$ & $1.05 $ & $0.0419$ & $150$ & $0.510$\\
\hline
\end{tabular}
\end{center}
\caption{Parameter values:  Origins of the values are described in the text.}
\label{tab:para}\end{table}

\begin{figure}
\includegraphics[width=6cm]{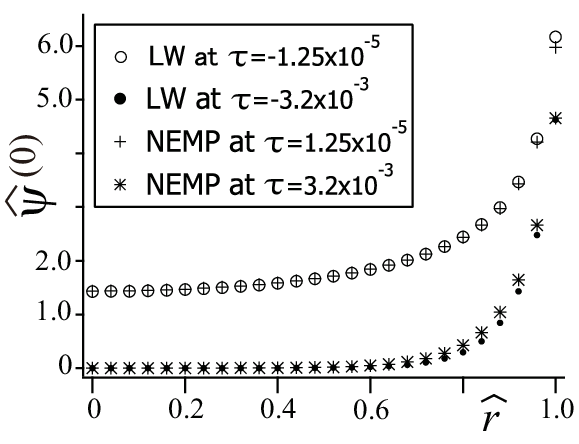}
\caption{Plots of the dimensionless order parameter at equilibrium, ${\hat\psi}^{(0)}({\hat r})$, against  
the dimensionless radial distance ${\hat r}$.
{The surface field is set to $h=0.1\ $cm$^3/$s$^2$.}
For a mixture of LW (NEMP), open circles (crosses) represent ${\hat\psi}^{(0)}({\hat r})$
at $|\tau|=1.25\times 10^{-5}$, whereas 
closed circles (asterisks) represent ${\hat\psi}^{(0)}({\hat r})$ 
at $|\tau|=3.2\times 10^{-3}$.  We use the values of $\tau_*$ in Table \ref{tab:para} to find
 ${\hat \tau}=-0.24\ (0.19)$ for open circles (crosses), and ${\hat \tau}=-63\ (49)$ for closed circles (asterisks).} 
\label{fig:profile}
\end{figure}

\subsection{Equilibrium profile and thermal force density}
Equilibrium profiles in the tube are shown   
in Fig.~\ref{fig:profile}.  Because {of ${\hat h}>0$}, ${\hat \psi}^{(0)}({\hat r})$ increases,
or the component a is more concentrated,
near the wall at ${\hat r}=1$.   For each mixture, ${\hat \psi}^{(0)}({\hat r})$ is larger at the smaller value
of $|\tau|$, since the adsorption layer extends towards the tube's center 
as the critical temperature is approached. 
{At $|\tau|=3.2\times 10^{-3}$,} {the adsorption layer appears to localize near the tube's wall, and}  
{${\hat \psi}^{(0)}({\hat r})$ is larger for a mixture of NEMP than for a mixture of LW
in the whole region of $0\le {\hat r}\le 1$ although the difference is hard to see for ${\hat r}<0.7$ and ${\hat r}=1$ in the figure.
This magnitude relationship is reasonable considering that $|{\hat \tau}|$ is smaller for a mixture of ${\rm NEMP}$.
At $|\tau|=1.25\times 10^{-5}$, the relationship {holds only for  ${\hat r}<0.7$, and 
$\xi$} at $\psi=0$, given by $\xi_0|\tau|^{-\nu}$, exceeds the tube radius.
It is approximately equal to $2 r_{\rm tube}$ at $|\tau|=1.25\times 10^{-5}$ while
to $r_{\rm tube}/10$ at $|\tau|=3.2\times 10^{-3}$ for both mixtures.
At ${\hat r}=0$ under $|\tau|=1.25\times 10^{-5}$, the composition is definitely off-critical
and $\xi$ becomes approximately equal to
$r_{\rm tube}/2$ for both mixtures.}  \\

\noindent
Circles in Fig.~\ref{fig:inte} represent  
${\hat \sigma}_z^{({\rm th})}$ of Eq.~(\ref{eqn:thermal-force-density0})
at $|\tau|=1.25\times 10^{-5}$.
Hereafter, $\tau$ ($\hat{\tau}$) represents the (scaled) reduced temperature in the reference state.
{Figures \ref{fig:inte}(a) and \ref{fig:inte}(b)  
show that the thermal force density becomes remarkable near the tube's wall.}
{The first term of Eq.~(\ref{eqn:hatf}), which
originates from $A_0\psi^2/2$ in the term of $\psi$ squared in the bare model, 
contributes to the second term on the RHS of Eq.~(\ref{eqn:thermal-force-density0}) via
the term $\partial{\hat f}/(\partial{\hat \tau})$.
{Crosses in Fig.~\ref{fig:inte} represent
this contribution, which} is denoted by ${\hat\sigma}^{({\rm sq})}$ and is given by
\begin{equation}
{\hat\sigma}_z^{({\rm sq})}({\hat r})=\pm\frac{T^{(0)}}{2 T_{\rm c}}{\hat\omega}^{\gamma-1}{\hat\psi}^2
\ .\label{eqn:domi}\end{equation}
This is evaluated at $\varepsilon=0$ with the same sign as $\tau$ being taken. {This sign  for 
${\hat \psi}\ne 0$ comes from that of $A_0$, which is negative (positive)
for the LC (UC) point.}
The rest in the second term on the RHS of Eq.~(\ref{eqn:thermal-force-density0})
is plotted with triangles.  
The first term on the RHS of Eq.~(\ref{eqn:thermal-force-density0}), which originates from 
the scalar-pressure deviation Eq.~(\ref{eqn:scalar-pressure}),
is plotted with squares.  This term gives negligibly small contributions 
to ${\hat \sigma}_z^{({\rm th})}$} {in the whole region of ${\hat r}$. It}
remains the case for  $|\tau|$ up to $6.4\times 10^{-3}$ although data are not shown.
This can be expected because the first term contains a small positive factor $\tau_*$.
{Near the wall in Fig.~\ref{fig:inte}, we can see that   
${\hat\sigma}_z^{({\rm sq})}({\hat r})$ is dominant in ${\hat\sigma}_z^{({\rm th})}({\hat r})$.  
{The ratio ${\hat\sigma}_z^{({\rm sq})}({\hat r})/{\hat\sigma}_z^{({\rm th})}({\hat r})$ at ${\hat r}=1$}  
 is $0.88\ (0.87)$ for $|\tau|=1.25\times 10^{-5}$ in a mixture of LW (NEMP).   At $|\tau|={3.2}\times 10^{-3}$, 
{the ratio at ${\hat r}=1$} remains approximately the same, $0.86\ (0.85)$, although each {of  
${\hat\sigma}_z^{({\rm sq})}({\hat r})$ and ${\hat\sigma}_z^{({\rm th})}({\hat r})$ at ${\hat r}=1$}  is roughly halved.}
\\

\begin{figure}
\includegraphics[width=12cm]{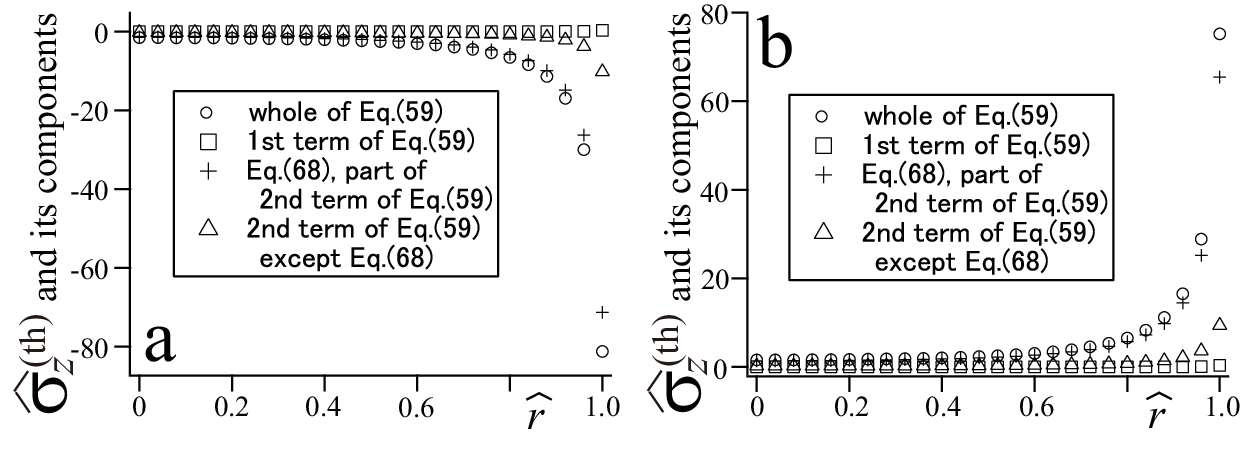}
\caption{Plots of the scaled thermal force density, ${\hat \sigma}_z^{({\rm th})}$,
and its components  
against the dimensionless radial distance ${\hat r}$ at $|\tau|=1.25\times 10^{-5}$ 
for a mixture of LW (a) and a mixture of NEMP (b). Here, the reduced temperature $\tau$ is evaluated at $T=T^{(0)}$, 
and is negative (positive) in the former (latter) mixture.  The surface field is set  to $h=0.1\ $cm$^3/$s$^2$.
Circles represent ${\hat \sigma}_z^{({\rm th})}$ of 
 Eq.~(\ref{eqn:thermal-force-density0}), whereas
squares represent its first term.  
Its second term can be separated into two parts; ${\hat\sigma}_z^{({\rm sq})}$ of Eq.~(\ref{eqn:domi}) and the rest.
Crosses represent the former, whereas triangles represent the latter.}
\label{fig:inte}
\end{figure}

\noindent 
{In Fig.~\ref{fig:inteh} for a mixture of NEMP, 
${\hat\sigma}_z^{({\rm sq})}({\hat r})$
and ${\hat\sigma}_z^{({\rm th})}({\hat r})$ become distinctly larger near the wall as $\tau$ is smaller and $h$ is larger}.
For various values of $|\tau|$ and $h$ examined in Figs.~\ref{fig:inte} and \ref{fig:inteh},
over the whole region of ${\hat r}$,
the ratio {${\hat\sigma}_z^{({\rm sq})}({\hat r})/{\hat\sigma}_z^{({\rm th})}({\hat r})$}
remains {approximately the same as the ratio at ${\hat r}=1$.  Thus,  
as far as examined}, over the whole region of ${\hat r}$,
${\hat \sigma}_z^{({\rm th})}({\hat r})$ is
negative (positive) in a mixture of LW (NEMP) and is contributed dominantly from
${\hat\sigma}_z^{({\rm sq})}({\hat r})$.
In each inset of Fig.~\ref{fig:inteh}, ${\hat \psi}^{(0)}({\hat r})$
increase more steeply near the wall as $h$ is larger, like
${\hat \sigma}_z^{({\rm th})}({\hat r})$ in the main figure.
For each value of $h$ {in Figs.~\ref{fig:inteh}(a) and \ref{fig:inteh}(b)},
{as ${\hat r}$ decreases, both ${\hat \psi}^{(0)}({\hat r})$ and ${\hat \sigma}_z^{({\rm th})}({\hat r})$
decrease  more gradually at the smaller value of $\tau$.   
These behaviors}
can be explained by the dominance of Eq.~(\ref{eqn:domi}).
The dominance of Eq.~(\ref{eqn:domi}) in the term involving $\partial{\hat f}/({\partial\tau})$ on the RHS of
Eq.~(\ref{eqn:thermal-force-density0}) 
is expected from approximate estimation mentioned below Eq.~(\ref{eqn:omegatau}).

\begin{figure}
\includegraphics[width=12cm]{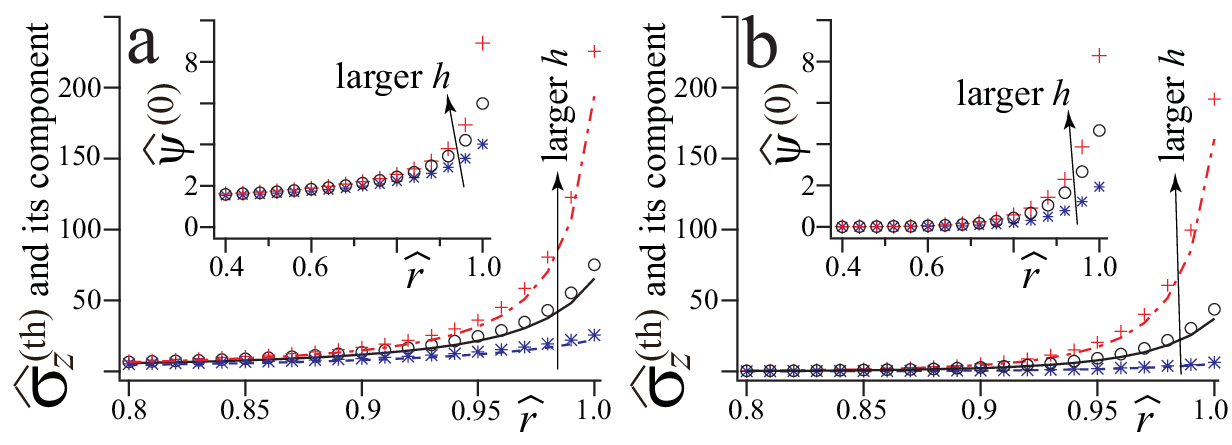}
\caption{Plots of the {scaled} thermal force density, ${\hat \sigma}_z^{({\rm th})}$,
and its dominant term ${\hat\sigma}_z^{({\rm sq})}$ 
 against the dimensionless radial distance ${\hat r} (\ge 0.8)$ for a mixture of ${\rm NEMP}$.   
The reduced temperature $\tau$ evaluated at $T=T^{(0)}$ is $1.25\times 10^{-5}$ in (a)
and $3.2\times 10^{-3}$ in (b).  The surface field $h$ is set to $10^{-1}\ $cm$^3/$s$^2$ for
circles (${\hat \sigma}_z^{({\rm th})}$) and solid curves (${\hat\sigma}_z^{({\rm sq})}$); 
these results in (a) are already shown in Fig.~\ref{fig:inte}(b).  Setting $h$ 
to $10^{-0.5}\ (10^{-1.5})\ $cm$^3/$s$^2$,
{we obtain results shown by red crosses and dash-dot curves 
(blue asterisks and dashed curves); symbols represent ${\hat \sigma}_z^{({\rm th})}$ and curves represent
${\hat\sigma}_z^{({\rm sq})}$}.
The change of $h$ is indicated by arrows.
(Insets) Plots of the dimensionless order parameter at equilibrium, $\hat{\psi}^{(0)}({\hat r})$, against ${\hat r}{(\ge 0.4)}$.
The parameter values for each symbol are the same as those for the same symbol in the main figure
in each of (a) and (b). {The results of the circles are already shown in Fig.~\ref{fig:profile}}.}
\label{fig:inteh}
\end{figure}

\subsection{Velocity field and conductance}
We define ${\hat v}_z^{({\rm th})}({\hat r})$ as the double integral of Eq.~(\ref{eqn:nodimv0}), 
which is plotted in Fig.~\ref{fig:vel}.  When $\delta T$ is positive,
${\hat v}_z^{({\rm th})}$ has the same sign as $v_z^{({\rm th})}$.
At $|\tau|=3.2\times 10^{-3}$, ${\hat v}_z^{({\rm th})}({\hat r})$ changes only for ${\hat r}>0.8$ and 
the velocity appears to slip across a narrow region near the wall.  This is because, as shown in Fig.~\ref{fig:inteh}(b), 
the adsorption layer and the thermal force density localize sharply in a region near the wall ${\hat r}>0.8$. 
{The slip velocity is given by
 ${\hat v}_z^{({\rm th})}(0)$, which is $-0.042\ (0.061)$ for a mixture of LW (NEMP) at $h=0.1\ $cm$^3/$s$^2$}.
{Converting the value to the slip velocity with dimensions, we find it to be} 
$-7.09\ (38.2)\ (\mu{\rm m})^2/($s$\cdot$K$)$ multiplied by $\delta T/L_{\rm tube}$, which is comparable in magnitude 
with typical thermophoretic mobility far from the critical point \cite{piazza, Jiang, Braun, maeda}. 
{For each value of $h$ in Fig.~\ref{fig:vel}, $|{\hat v}_z^{({\rm th})}({\hat r})|$}  
at $|\tau|=1.25\times 10^{-5}$ is larger than $|{\hat v}_z^{({\rm th})}({\hat r})|$  
at $|\tau|=3.2\times 10^{-3}$ inside the tube and increases gradually in magnitude as ${\hat r}$ decreases
without showing an obvious slip. 
{In Fig.~\ref{fig:vel}(b), ${\hat v}_z^{({\rm th})}({\hat r})$ increases with $h$ inside the tube, as expected}.
The spatial resolution of our formulation is given by $\xi$.
For $h=0.1\ $cm$^3/$s$^2$ and $|\tau|=1.25\times 10^{-5}\ (3.2\times 10^{-3})$,
a mixture of LW has $\xi/r_{\rm tube}=0.030\ (0.036)$ 
and a mixture of NEMP has $0.032\ (0.038)$ at ${\hat r}=1$.
With the spatial resolution given by these values, one would trace rapid changes of ${\hat v}_z^{({\rm th})}$ near the wall
shown in Fig.~\ref{fig:vel}.  In passing, the slip velocity at $|\tau|=3.2\times 10^{-3}$ can be evaluated
approximately using the Gaussian model mentioned in Appendix \ref{sec:gauss}\@.\\

\begin{figure}
\includegraphics[width=12cm]{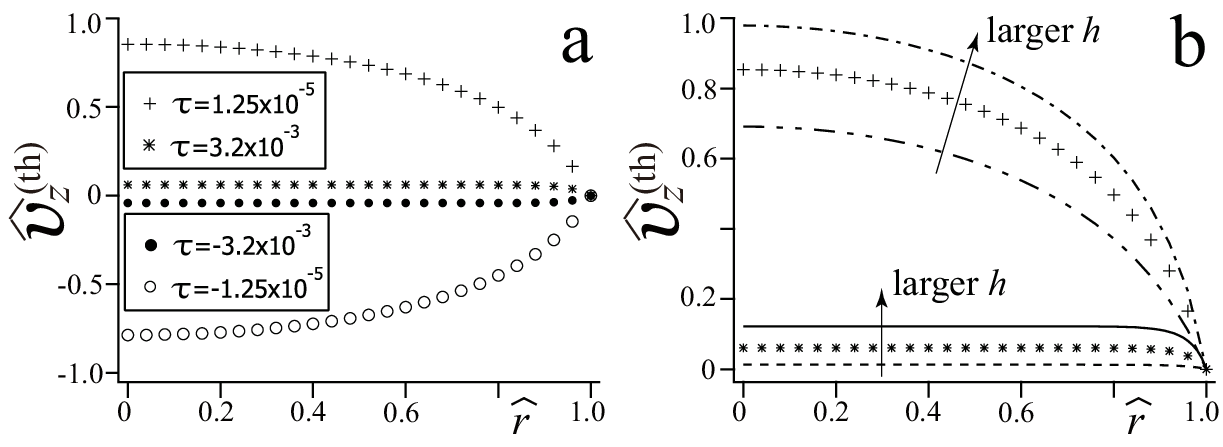}
\caption{The $z$ component of the dimensionless velocity in thermoosmosis, ${\hat v}_z^{({\rm th})}({\hat r})$,
is plotted against the dimensionless radial distance ${\hat r}$.  
(a) Closed and open circles represent ${\hat v}_z^{({\rm th})}({\hat r})$ for a mixture of LW
at $\tau =-3.2\times 10^{-3}$ and $-1.25\times 10^{-5}$, respectively.
The reduced temperature $\tau$ is evaluated at $T=T^{(0)}$.
Asterisks and crosses represent ${\hat v}_z^{({\rm th})}({\hat r})$ for a mixture of NEMP
at $\tau =3.2\times10^{-3}$ and $1.25\times 10^{-5}$, respectively. The 
surface field is set to $h=10^{-1}\ $cm$^3/$s$^2$. 
{(b) Asterisks and crosses represent the same results as those in (a), respectively.
The solid curve (the dashed curve)} represents ${\hat v}_z^{({\rm th})}({\hat r})$ for a mixture of NEMP at  $\tau =3.2\times10^{-3}$ 
with $h$ set  to $10^{-0.5}\ (10^{-1.5})\ $cm$^3/$s$^2$.  The dash-dot curve 
(the dash-dot-dot curve) 
represents ${\hat v}_z^{({\rm th})}({\hat r})$ at 
$\tau=1.25\times 10^{-5}$ with $h$ set to $10^{-0.5}\ (10^{-1.5})\ $cm$^3/$s$^2$.  
The change of $h$ is indicated by arrows.}
\label{fig:vel}
\end{figure}

\noindent
The dimensionless thermoosmotic conductance is defined at Eq.~(\ref{eqn:conduct}), which is rewritten as
\begin{equation}
{\hat G}^{({\rm th})}=16\int_0^{1}d{\hat r}\ {\hat r}{\hat v}_z^{({\rm th})}({\hat r})\ .\label{eqn:conductb}
\end{equation}
When $\delta T$ is positive, ${\sigma}_z^{({\rm th})}{({\hat r})}$ has the same sign 
as ${\hat\sigma}_z^{({\rm th})}{({\hat r})}$.
For each mixture in our numerical results, the sign of
${\hat\sigma}_z^{({\rm th})}({\hat r})$ remains the same for $0\le{\hat r}\le 1$,
and thus is the same as that of
${\hat v}_z^{({\rm th})}({\hat r})$ and that of ${\hat G}^{({\rm th})}$;
${\hat G}^{({\rm th})}>0\ (<0)$ means that the flow
direction is the same as (opposite to) {the direction of} the temperature gradient.
{Thus, according to our numerical results}, a mixture of NEMP near the UC point flows 
towards the reservoir with the higher temperature, whereas
a mixture of LW near the LC point flows in the opposite direction.  {This is}
independent of which component is adsorbed onto the tube's {wall}
{owing to the independence of Eq.~(\ref{eqn:thermal-force-density0}) from the sign of $h$}.  
\\

\begin{figure}
\includegraphics[width=6cm]{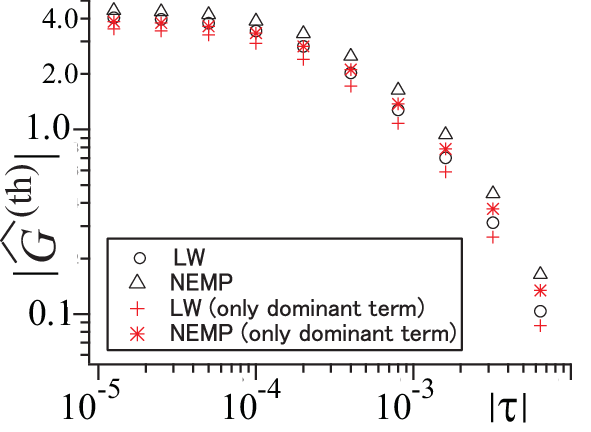}
\caption{Logarithmic plots of the absolute value of the dimensionless  thermoosmotic conductance
against that of the reduced temperature $|\tau|$, evaluated at $T=T^{(0)}$, for a mixture of LW (circle) and a mixture of 
NEMP (triangle) with the surface field set to $h=0.1\ $cm$^3/$s$^2$. 
The {dimensionless} conductance, ${\hat G}^{({\rm th})}{=\Omega[1, {\hat \sigma}_z^{({\rm th})}]}$ 
of Eq.~(\ref{eqn:conduct}) or (\ref{eqn:conductb}),
is negative for a mixture of LW, whereas
positive for a mixture of NEMP\@. {Red crosses (asterisks) represent $\Omega[1, {\hat \sigma}_z^{({\rm sq})}]$, which is}
{dominant in ${\hat G}^{({\rm th})}$},  
{for a mixture of LW (NEMP)}.   
}
\label{fig:massfl}
\end{figure}

\noindent
Logarithmic plots of $|{\hat G}^{({\rm th})}|$
against $|\tau|$ are shown in Fig.~\ref{fig:massfl}, {where}
the conductance increases in magnitude as $|\tau|$ decreases.
This is because larger susceptibility makes the PA stronger.
For smaller values of  $|\tau|$, the increase becomes more gradual.  This 
would represent effect of the size of the tube, {considering that
the value of $\xi$ at $\psi=0$, given by $\xi_0|\tau|^{-\nu}$,
exceeds the tube radius approximately for $|\tau|<5\times 10^{-5}$.
 Equation (\ref{eqn:domi}), ${\hat \sigma}_z^{({\rm sq})}$, contributes to
${\hat G}^{({\rm th})}$ dominantly in the range of $\tau$ examined in Fig.~\ref{fig:massfl}}.  
{Changing the value of $h$ for a mixture of NEMP, 
we calculate the conductance,  
as shown in Fig.~\ref{fig:hatGh}.   
As $h$ increases, {${\hat G}^{({\rm th})}$ increases, as expected 
since ${\hat \psi}^{(0)}$, ${\hat \sigma}_z^{({\rm th})}$,  and ${\hat v}_z^{({\rm th})}$ then increase
in Figs.~\ref{fig:inteh} and \ref{fig:vel}(b).}
In Fig.~\ref{fig:hatGh}(b), {${\hat G}^{({\rm th})}$ becomes less dependent} on $h$ in logarithmic scale 
as $\tau$ decreases.
This tendency is also observed {for the dependence of ${\hat v}_z^{({\rm th})}$ on $h$} in Fig.~\ref{fig:vel}(b).
{The contribution from ${\hat \sigma}_z^{({\rm sq})}$ to
${\hat G}^{\rm (th)}$ remains dominant}  for the values of $\tau$ and ${\hat h}$ 
examined in Fig.~\ref{fig:hatGh}(b).}

\begin{figure}
\includegraphics[width=12cm]{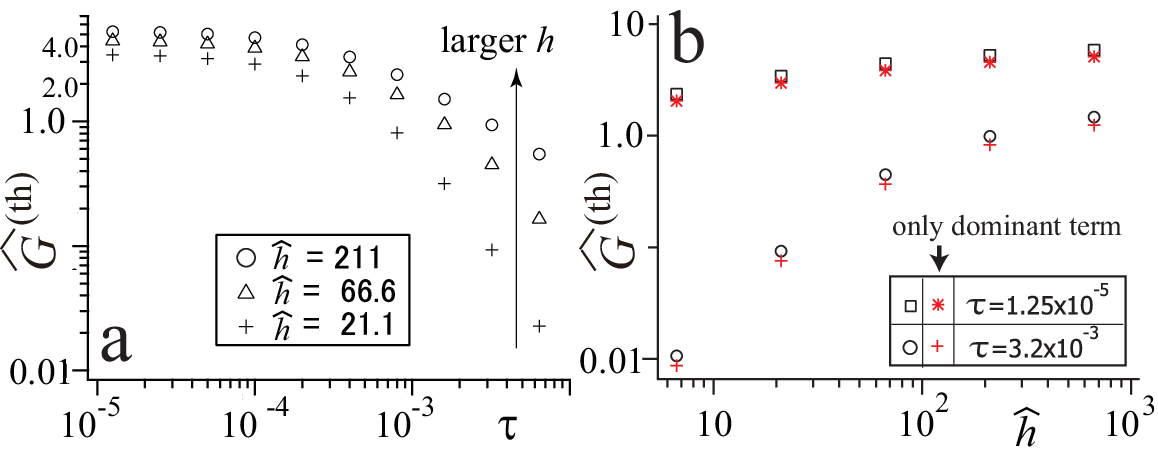}
\caption{(a) Logarithmic plots of the dimensionless thermoosmotic conductance, ${\hat G}^{({\rm th})}$
against the reduced temperature $\tau$, evaluated at $T=T^{(0)}$, for a mixture of NEMP\@.
Triangles in this figure and in Fig.~\ref{fig:massfl} represent the same results {with the surface field
set to $h=10^{-1}\ $cm$^3$/s$^2$.
Circles (crosses) represent}
${\hat G}^{({\rm th})}$ for $h$=$10^{-0.5}\ (10^{-1.5})\ $cm$^3/$s$^2$. 
{The corresponding values of the scaled surface field ${\hat h}$ are shown in the figure;
the} change of $h$ is indicated by an arrow.
(b) Logarithmic plots of ${\hat G}^{({\rm th})}$
against the scaled surface field ${\hat h}$ for a mixture of NEMP\@.
Squares and circles represent ${\hat G}^{({\rm th})}=\Omega[1, {\hat \sigma}_z^{({\rm th})}]$ 
at $\tau=1.25\times 10^{-5}$ and $3.2\times 10^{-3}$,
respectively. The values of $\tau$ are evaluated at $T=T^{(0)}$.  
Red asterisks (crosses) represent
$\Omega[1, {\hat \sigma}_z^{({\rm sq})}]$, which is dominant in ${\hat G}^{({\rm th})}$, 
at $\tau=1.25\times 10^{-5}$ ($3.2\times 10^{-3}$). The change of $\tau$ is indicated by an arrow.}
\label{fig:hatGh}
\end{figure}

\subsection{Prediction of universal properties}
{As mentioned below Eq.~(\ref{eqn:domi}), in our numerical results,
the first term on the RHS of Eq.~(\ref{eqn:thermal-force-density0}) is negligible.
This} {would be mainly because it contains a very small positive factor $\tau_*$ (Table \ref{tab:para}). 
Thus, owing to $T_{\rm c}\approx T^{(0)}$, it is strongly suggested that
\begin{equation}
{\hat\sigma}_z^{({\rm th})}
\approx \frac{\partial {\hat f}}{\partial{\hat \tau}}-\frac{\eta \nu}{2{\hat \omega}^{\eta\nu+1}}
{\frac{\partial{\hat\omega}}{\partial{\hat \tau}}}\left|\partial_{\hat r} {\hat\psi}\right|^2
\ ,\label{eqn:thermalapp1}
\end{equation}
which is evaluated at $\varepsilon=0$, holds for any mixture.
The RHS of Eq.~(\ref{eqn:thermalapp1}) is free from the material constants, {\it i.e.\/}
does not suppose a special mixture, because
it is determined only by the scaled reduced-temperature ${\hat \tau}$ 
and the magnitude of the scaled surface field
$|{\hat h}|$.
Using various values of $({\hat \tau}, {\hat h})$, we numerically find that 
Eq.~(\ref{eqn:domi}) is dominant in ${\hat \sigma}_z^{({\rm th})}$.
In this sense, we have
\begin{equation}
{\hat\sigma}_z^{({\rm th})}\approx 
\pm\frac{{\hat\omega}^{\gamma-1}}{2}{\hat\psi}^2
\label{eqn:thermalapp2}
\end{equation}
evaluated at 
$\varepsilon=0$, where the sign is taken as
that of $\tau$, {\it i.e.\/}, as that of $A_0$ in the bare model. 
Notably, Eq.~(\ref{eqn:thermalapp2}) is strongly expected to hold in the critical regime for any mixture,
which is also supported by the approximate estimation given below Eq.~(\ref{eqn:omegatau})
and  by the results in Figs.~\ref{fig:massfl} and \ref{fig:hatGh}(b).
Therefore, we can predict that}, 
{for any mixture near the UC (LC) point},  
the direction of thermoosmosis
{is the same as (opposite to) that of the temperature gradient,
irrespective of which component is adsorbed onto the wall}, {if
the critical composition is kept in the middle of each reservoir}.

\section{Further discussion and summary\label{sec:disc}}
{Our numerical results are based on the calculation up to the order of $\varepsilon$,
or in the linear regime with respect to $\delta T$}.
Obviously, after a temperature difference $\delta T$ is imposed between the reservoirs, 
the deviation of $\tau$ from the value of $\tau$ at $T=T^{(0)}$ is required to be much smaller 
in magnitude than the value at $T=T^{(0)}$ throughout inside the tube.  For example,
if the latter value is set to $10^{-3}$ in magnitude,
we may set $|\delta T|/T_{\rm c}$ to be smaller than its $10\ \%$, {$10^{-4}$.
The resultant local changes of $T$ and ${\hat \psi}$ shift ${\hat \sigma}_z^{({\rm th})}$.
As far as examined, the shift is {roughly smaller than $10\ \%$ in the adsorption layer}.  
For $|\delta T|=100\ {\rm mK} \ll |T^{(0)}-T_{\rm c}|\approx 1\ $K and $L_{\rm tube}=10\ \mu$m},
we find from the results in  Fig.~\ref{fig:vel}(a) that the slip velocity is approximately $0.1\ \mu$m$/$s, 
which would be measured experimentally.   In passing, in the experiments on
the Brownian motion of colloidal particles in a mixture, $|\tau|$ is set to be smaller than $10^{-4}$ 
homogeneously \cite{omari, beys2019}.
\\

\noindent
The RLFT succeeds in describing several phenomena of a mixture \cite{rlft}. 
However, in the theory, crossover to the regular part of the free energy \cite{wegner, CAS, CATS, edison, vanthof, olaya}
is not considered, the results up to the one-loop order approximation are used, and   
validity of the definition of the local  correlation length in the inhomogeneous
composition is not fully discussed.   The regular parts of the transport coefficients \cite{folkmos2} are 
considered only for the viscosity $\eta_{\rm s}$, whose singularity is very weak, in the present study.
These points are to be improved in future 
for quantitatively better {numerical results for transport properties}, {not only of
thermoosmosis but also of} {various phenomena, such as cross effects, described by the Onsager coefficients.}
Still, the qualitative property on the flow direction {in thermoosmosis of a mixture}, {predicted by the present study},
should be robust to changes of details in the formulation, considering that 
it originates from the sign of the coefficient, $A_0$, in the bare model. \\

\noindent
Equation (\ref{eqn:thermalapp2}) leads to
a possible power-law dependence with respect to $|\tau|$
for the slip velocity far from a flat wall in thermoosmosis, as shown in Appendix \ref{sec:gauss}\@.
{Thermophoresis would occur for a particle 
in a mixture in the presence of PA onto the particle surface;
the direction of the particle motion is expected to be the same as (opposite to)
that of the imposed temperature gradient if the mixture is near the LC (UC) point 
and has the the critical composition far from the particle.}
These points clearly require further investigation in future.\\

\noindent
{Our present study is summarized as follows.} 
We consider transport of a binary fluid mixture, lying in the one-phase region near the demixing critical point, 
through a capillary tube.
One component is assumed to be adsorbed onto the tube's adiabatic wall and
{the adsorption layer can be much thicker than} the molecular size. 
We formulate the hydrodynamics from
a coarse-grained free-energy functional using an extended Gibbs-Duhem relation, Eq.~(\ref{eqn:nablaPi}), 
{consistently with principles of linear nonequilibrium thermodynamics}.
This relation is originally derived in Ref.~\cite{dvw}, and is derived on a more general condition in Appendix \ref{sec:stress}\@. 
{Assuming the critical composition in} the middle of each reservoir in the reference equilibrium state, 
we derive the Onsager coefficients in Section \ref{sec:formulae}.  
{Among various phenomena described by the coefficients}, we focus on thermoosmosis of a mixture.
{The formula for the thermal force density, Eq.~(\ref{eqn:thermal-force-density}), is explicitly derived 
as an extension of Derjaguin and Sidorenkov's
formula for one-component fluids
and is rewritten as Eq.~(\ref{eqn:thermal-force-density0}) in terms of the renormalized local functional theory \cite{fisher-degennes,rlft}.}
We predict that
the direction of thermoosmotic flow of a mixture near the upper (lower) consolute point is the same as (opposite to) 
that of the temperature gradient, irrespective of which component is adsorbed onto the tube's wall. 
The magnitude of the thermoosmotic conductance increases, with the increase being more gradual owing to the size effect, 
as the critical point is approached.
The thermal force density is given in a scaled form by Eq.~(\ref{eqn:thermalapp1}),
which depends only on the scaled reduced-temperature and the scaled surface field, {and is 
dominantly contributed from Eq.~(\ref{eqn:thermalapp2})}.
\\

\noindent
Such mesoscopic inhomogeneity as is generated in a mixture by the surface field
can occur in many soft matter systems --- polymer solutions, polyelectrolytes, and liquid crystals \cite{onukibook, degennes}.
In particlular, their dynamics driven by a temperature gradient would be studied by applying our procedure to
a suitable set of hydrodynamic equations based on a coarse-grained free-energy functional. 
Also, for thermoosmosis of a solution far from the critical point,
our results may help as a guide regarding properties independent of the microscopic details.   
Hence, our present study would lay solid foundations on non-isothermal hydrodynamics in the presence of mesoscopic inhomogeneity
and predict universal properties on thermoosmosis of a near-critical binary fluid mixture. 

\appendix
\begin{section}{Non-dissipative part of the stress tensor\label{sec:stress}}
For conciseness, we here write $(T, \rho_n, \nabla \rho_n)$ for the variables of $f_{\rm bulk}$ in Eq.~(\ref{eqn:general}).
They are also variables of $s$ and $u$ because of Eq.~(\ref{eqn:sandu}).
The entropy density is also a function of $u$, $\rho_n$, and $\nabla\rho_n$, and we define ${\tilde s}$ so that
\begin{equation}
s(T, \rho_n, \nabla\rho_n)={\tilde s}(u(T, \rho_n, \nabla\rho_n), \rho_n,
\nabla\rho_n)\label{eqn:ent}
\end{equation}   
holds. Explicit expressions of ${\tilde s}$, although 
given in special cases \cite{dvw,gonn}, are not required 
{in a general argument given below}.
We have
\begin{equation}
\frac{\partial {\tilde s}}{\partial u}=\frac{1}{T}\quad {\rm and}\quad 
\frac{\partial {\tilde s}}{\partial \zeta}
=-\frac{1}{T} \frac{\partial {f_{\rm bulk}}}{\partial \zeta}  \label{eqn:rensa}
\end{equation}
for ${\zeta}=\rho_n$ or $\nabla\rho_n$.  {As mentioned above Eq.~(\ref{eqn:general})},
we assume the dependence of $f_{\rm bulk}$ on $\nabla\rho_n$
to be through a quadratic form, whose coefficients may depend on $T$ and $\rho_n$.
Below, as in Appendix A of Ref.~\cite{pipe},
we consider a quasistatic deformation of a mixture to derive Eqs.~(\ref{eqn:chempot}) and (\ref{eqn:Pidef}).
We write $V_t$ for a small region co-moving with the deformation.
Here, $t$ is not time but a parameter of the deformation. In general, 
an infinitesimal change in the entropy 
are contributed independently from the mechanical work, from the change in
the composition, and from the change in the internal energy.
Thus, regarding $T$, ${\mathsf \Pi}$, and $\mu_n$ as homogeneous
over a small region $V_t$, we have
\begin{equation}
T\frac{d}{dt}\int_{V_t} d{\v r}\ {\tilde s}={\mathsf \Pi}: \int_{\partial V_t} dA\ {\v n}_{\partial V_t}{\v v}
-\mu_n\ \frac{d}{dt}\int_{V_t} d{\v r}\ \rho_n({\v r}, t)+\frac{d}{dt} \int_{V_t}d{\v r}\ u
\ .\label{eqn:loc}\end{equation}
{Here}, the symbol $:$ is defined so that ${\mathsf A}:{\mathsf B}=
{\mathsf A}_{ij}{\mathsf B}_{ji}$ holds for two tensors ${\mathsf A}$ and ${\mathsf B}$,
{and ${\v n}_{\partial V_t}$ is the outward facing unit normal vector of the surface of $V_t$}. \\

\noindent
Each locus of a mixture is assumed to have each bath of particles and heat.
We here write ${\v j}_n$ and ${\v j}_u$ for their respective fluxes 
to the bath, and write ${\v v}$ for a displacement vector per unit value of $t$.
The meanings of ${\v j}_n$ and ${\v v}$ are different from the ones in the text, respectively;
${\v j}_{\rm a} +{\v j}_{\rm b}$ does not always vanish here.
Although $t$ is not the time, 
we can treat $t$ as the time formally to define the Eulerian time-derivative $\partial/(\partial t)$ and
Lagrangian time-derivative $D/(Dt)$.
We have
\begin{equation}
\frac{D\rho_n}{Dt}=-\rho_n\nabla\cdot {\v v} -\nabla\cdot {\v j}_n\quad {\rm and}\quad 
\frac{Du}{Dt}=-u\nabla\cdot {\v v} -\nabla\cdot {\v j}_{\rm u}
\ .\label{eqn:revhozon}\end{equation}
The whole region occupied by the mixture, $V_{\rm tot}$, is deformable here, unlike in the text.  
The LHS of Eq.~(\ref{eqn:loc}) is rewritten as the integral of $T[D{\tilde s}/(Dt)+{\tilde s}\nabla\cdot {\v v}]$
over $V_t$.  Rewriting {the last two} terms on the RHS similarly
{and applying the divergence theorem for the first term}, we obtain an equation for the integrands owing to arbitrariness
of $V_t$.  With the aid of this equation,  
the change in the entropy in {$V_{\rm tot}$} per unit value of $t$ is found to be
\begin{equation}
\int_{V_{\rm tot}} d{\v r}\ \left[\frac{D{\tilde s}}{Dt}+{\tilde s}\nabla\cdot {\v v}\right]
=\int_{V_{\rm tot}}d{\v r}\ \left[\frac{{\mathsf \Pi}}{T}: \nabla {\v v}+\frac{\mu_n}{T}\nabla \cdot {\v j}_n -\frac{1}{T}\nabla\cdot{\v j}_{\rm u}
\right]
\label{eqn:Shenka}\ ,\end{equation} 
where $T$, ${\mathsf \Pi}$, and $\mu_n$ can be inhomogeneous.  
The factor $\nabla\cdot {\v v}$ in Eqs.~(\ref{eqn:revhozon}) and (\ref{eqn:Shenka}) comes from
the change rate of the Jacobian between the Eulerian and Lagrangian coordinates. 
We have
\begin{equation}
\frac{D{\tilde s}}{Dt}=\frac{\partial {\tilde s}}{\partial u}\frac{Du}{Dt}+\frac{\partial {\tilde s}}{\partial \rho_n} \frac{D\rho_n}{Dt}+
\frac{\partial {\tilde s}}{\partial \left(\nabla\rho_n\right)}\cdot \nabla\left(\frac{D\rho_n}{Dt}\right)
-\frac{\partial {\tilde s}}{\partial \left(\nabla\rho_n\right)}\cdot \left(\nabla{\v v}\right)\cdot\left(\nabla\rho_n\right) 
\ ,\label{eqn:dsdt}\end{equation}
which can be rewritten using Eqs.~(\ref{eqn:rensa}) and (\ref{eqn:revhozon}).
Substituting the result into the LHS of Eq.~(\ref{eqn:Shenka}) and applying
integration by parts, we find the LHS to be the sum of
\begin{equation}
-\int_{\partial V_{\rm tot}}d{\v r}\ \frac{1}{T}\frac{D\rho_n}{Dt}\frac{\partial f_{\rm bulk}}{\partial (\nabla\rho_n)}\cdot {\v n}_{\partial V_{\rm tot}}
\label{eqn:hyoumen}\end{equation}
and the RHS of Eq.~(\ref{eqn:Shenka}) with $\mu_n$ and ${\mathsf \Pi}$ being replaced by the RHS's of
Eqs.~(\ref{eqn:chempot}) and (\ref{eqn:Pidef}), respectively.  This means that $\mu_n$ and ${\mathsf \Pi}$
are given by these equations, respectively. 
Because ${\mathsf \Pi}$ is symmetric, {we can derive Eq.~(\ref{eqn:nablaPi}), or equivalently, 
$\nabla\cdot({\mathsf \Pi}/T)=\rho_n\nabla ({\mu_n}/{T}) 
-u\nabla(1/{T})$, which is of the same form as Eq.~(2.44) of Ref.~\cite{dvw}}.
Equation (\ref{eqn:chempot}) can be used 
in calculating {not only $\varphi^{(0)}$ but also $\varphi^{(1)}$};
the latter need not be obtained in the present study.
\\

\noindent
We next consider thermodynamics of the mixture in a thin interfacial region {regarded as}
$\partial V_{\rm tot}$.
The free energy per unit area of this region is denoted by $f_{\rm surf}$ in Eq.~(\ref{eqn:general}),
and is here denoted by $f^{({\rm s})}$.  The superscript $^{({\rm s})}$ in general indicates a thermodynamic quantity
in $\partial V_{\rm tot}$; {a density with the superscript represents a quantity per unit area.}  
As in Eq.~(\ref{eqn:sandu}), we can 
introduce internal energy {${u}^{({\rm s})}$} and 
entropy ${s}^{({\rm s})}$ from $f^{({\rm s})}$.  These three quantities are functions of $T^{({\rm s})}$ and $\rho_n^{({\rm s})}$.
As Eq.~(\ref{eqn:loc}) yields Eq.~(\ref{eqn:Shenka}), 
an equation for a small co-moving area on $\partial V_{\rm tot}$
yields an equation representing the change of the entropy on $\partial V_{\rm tot}$.
Two points are to be noted in this derivation.
First, the mechanical contribution 
consists of a term involving the two-dimensional pressure tensor ${\mathsf \Pi}^{({\rm s})}$ and
a term involving the force normal to the small area.  
{Because $f^{({\rm s})}$ includes no gradients of 
mass densities, the pressure tensor is written as the two-dimensional scalar pressure
multiplied by the identity tensor on $\partial V_{\rm tot}$. The scalar pressure denoted by $P^{({\rm s})}$.
We define $P_{\rm n}^{({\rm s})}$ so that
the normal force is  $P_{\rm n}^{({\rm s})} {{\v n}_{\partial V_{\rm tot}}}$ per unit area.  }
Second, the factor coming from the change rate of the Jacobian is not $\nabla\cdot {\v v}$,
{appearing Eqs.~(\ref{eqn:revhozon}) and (\ref{eqn:Shenka})}, but
$\nabla_\parallel \cdot {\v v}_\parallel -2H_{\rm m} {\v v}\cdot{\v n}_{\partial V_{\rm tot}}$, where
${\v v}_\parallel$ is the projection of ${\v v}$ on the plane tangential to $\partial V_{\rm tot}$, 
$\nabla_\parallel \cdot {\v v}_\parallel$ indicates the divergence defined on 
$\partial V_{\rm tot}$, and
$H_{\rm m}$ denotes the mean curvature of $\partial V_{\rm tot}$ \cite{physica}.  The curvature is defined so that
it is positive when the center of curvature lies on the side directed by ${\v n}_{\partial V_{\rm tot}}$. \\

\noindent
The temperature at a local area on $\partial V_{\rm tot}$, $T^{({\rm s})}$,
should be equal to $T$ at its adjacent local region of $V_{\rm tot}$.  Similarly,
$\rho_\alpha^{({\rm s})}$ is determined by $\rho_\alpha$ at the adjacent region 
so that the former equals the latter multiplied by the interfacial region's width.
No other factor {is involved in determining} $\rho_\alpha^{({\rm s})}$, which means $\mu_\alpha^{({\rm s})}=0$.
Through these relationships, $f_{\rm surf}(T,\rho_n)$ equals $f^{({\rm s})}(T^{({\rm s})}, \rho_n^{({\rm s})})$. 
Taking Eq.~(\ref{eqn:hyoumen}) into account, we find 
\begin{equation} 
0=\frac{\partial f_{\rm surf}}{\partial\rho_n}+\frac{\partial f_{\rm bulk}}{\partial \left(\nabla\rho_n\right)}\cdot
 {\v n}_{\partial V}\quad {\rm at}\quad \partial V\label{eqn:bc}
\end{equation}
{from the equation representing the change of the entropy on $\partial V_{\rm tot}$.}
We also find {$P^{({\rm s})}=-f_{\rm surf}$, 
which} gives the Laplace pressure ${P_{\rm n}^{({\rm s})}}=-2f_{\rm surf}H_m$ \cite{effvis}.
Notably, $f_{\rm surf}$ equals the grand-potential density of $\partial V_{\rm tot}$ owing to $\mu^{({\rm s})}_n=0$.
{We need not 
consider these interfacial forces and
the force exerted on the mixture by the tube's wall 
in calculating the velocity field in the tube because the latter force is determined so that 
the no-slip condition is realized.}  
\end{section}

\begin{section}{Reciprocal relations\label{sec:incep}}
We consider two sets of flow fields, each being driven by
the thermodynamic forces
$\left(-\delta (P/T)_k, -\delta(\mu_-/T)_k, \delta(1/T)_k \right)$, with $k$ being i or ii. 
The resultant thermodynamic fluxes
and fields in the tube are also indicated by the subscript $_k$. 
Different ways of applying the divergence theorem
to the volume integral of ${\eta_{\rm s}^{(0)}} E^{(1)}_{{\rm i}} : E^{(1)}_{\rm ii}$ over the tube interior, denoted by $V_{\rm tube}$, give
\begin{equation}
\int_{V_{\rm tube}}d{\v r}\ {\v v}_{\rm ii}^{(1)} \cdot \left[\nabla\cdot\left({\eta_{\rm s}^{(0)}} E_{\rm i}^{(1)}\right)\right]
=\int_{V_{\rm tube}}d{\v r}\ {\v v}_{\rm i}^{(1)} \cdot \left[\nabla\cdot\left({\eta_{\rm s}^{(0)}} E_{\rm ii}^{(1)}\right)\right]
\label{eqn:EE}\end{equation}
with the aid of
Eq.~(\ref{eqn:nablavtube}) and the no-slip condition at the tube's wall.  Here, we
neglect effects of tube's edges on the laminar flow.
Substituting Eq.~(\ref{eqn:st3xy+}) into Eq.~(\ref{eqn:EE}), we find that  
\begin{eqnarray}
&&\int_{S_{tube}}dA\ v_{{\rm ii}, z}^{(1)}
\left[\frac{\rho^{(0)}}{\rho^{({\rm ref})}}\delta\left(\frac{P}{T}\right)_{{\rm i}}+
\left(\varphi^{(0)}-\frac{\varphi^{({\rm ref})}}{\rho^{({\rm ref})}}\right)\delta\left(\frac{\mu_-}{T}\right)_{\rm i}\right.\nonumber\\
&&\qquad\qquad \left. -\left(P^{(0)}+e^{(0)}-\frac{\rho^{(0)}e^{({\rm ref})}}{\rho^{({\rm ref})}}\right)\delta\left(\frac{1}{T}\right)_{{\rm i}}\right]
\end{eqnarray}
equals the above equation with the subscripts $_{\rm i}$ and $_{\rm ii}$ exchanged.
Putting $\delta (P/T)_{\rm i}$,  $\delta (\mu_-/T)_{\rm i}$, $\delta (P/T)_{\rm ii}$, and $\delta(1/T)_{\rm ii}$ equal to zero, 
we use Eqs.~(\ref{eqn:jzprofile}), (\ref{eqn:jqzprofile}), (\ref{eqn:calIcalJ}), and (\ref{eqn:calK}) to find
$L_{23}=L_{32}$.
Likewise, we can obtain $L_{13}=L_{31}$ by putting 
$\delta (P/T)_{\rm i}$,  $\delta (\mu_-/T)_{\rm i}$, $\delta (\mu_-/T)_{\rm ii}$, and $\delta(1/T)_{\rm ii}$ equal to zero.
The other reciprocal relations can be derived {similarly, as} shown in Appendix B of {Ref.~\cite{pipe}}. 
\end{section}

\begin{section}{Dissipative fluxes\label{sec:diss}}
In an equilibrium mixture, we consider a region where the mass densities are homogeneous.  
There, $f_{\rm bulk}$ is a function of $T$, $\rho$, and $\varphi$, 
and we have
\begin{equation}
\left.\frac{\partial \mu_-}{\partial \varphi}\right)_{T,P}
=\left.\frac{\partial \mu_-}{\partial \varphi}\right)_{T,\rho}+
\left.\frac{\partial \mu_-}{\partial \rho}\right)_{T,\varphi}
\left.\frac{\partial \rho}{\partial \varphi}\right)_{T,P}
\ .\label{eqn:diss1}\end{equation}
The first partial derivative of the second term on the RHS above equals
$\partial^2 f_{\rm bulk}/(\partial \rho \partial \varphi)$, which vanishes
because Eq.~(\ref{eqn:efu}) is assumed.  
The second derivative does not diverge, as mentioned in Appendix \ref{sec:fbulk}\@.  Thus, whether
$T$ and $P$ are fixed or $T$ and $\rho$ fixed, $\partial \mu_-/
(\partial \varphi)$ are the same and are regarded as equal to
the inverse of $\chi$, which appears in Eq.~(\ref{eqn:Lambda0}).
We have
\begin{equation}
\frac{1}{\chi}=\left.\frac{\partial \mu_-}{\partial c_{\rm a}}\right)_{T, P} \left.\frac{\partial c_{\rm a}}{\partial \varphi}\right)_{T, P} 
=\frac{1}{2\rho^2{\bar v}_+}\left.\frac{\partial \mu}{\partial c_{\rm a}}\right)_{T, P}\ ,  
\end{equation}
where ${\bar v}_+$ denotes $({\bar v}_{\rm a}+{\bar v}_{\rm b})/2$. 
The second equality above comes from 
Eq.~(34) of Ref.~\cite{pipe}.
Because a mixture we consider
has $\rho {\bar v}_+\approx 1$ {\cite{pipe}}, 
Eq.~(\ref{eqn:Lambda0}) is consistent with the result in Refs.~\cite{mist, luet}.\\

\noindent
With ${\bar \delta}$ indicating the deviation from the average,
the thermodynamic forces are ${\bar \delta}(1/T)$, $-{\bar\delta}(\mu_+/T)$, and $-{\bar\delta}(\mu_-/T)$
in Eq.~(\ref{eqn:linear1}). This equation is rewritten as
\begin{equation}
{\v j}=-4{\tilde \alpha}\nabla\mu_-+2{\tilde \beta}\nabla T\quad {\rm and}\quad 
{\v j}_{\rm q}-\mu_-{\v j}=2T {\tilde \beta}\nabla \mu_--{\tilde \gamma}\nabla T
\ ,\label{eqn:linear2}\end{equation}
whereby ${\tilde \alpha}$, ${\tilde \beta}$, and ${\tilde \gamma}$
are defined. We write ${\check s}(\equiv s/\rho)$ for
entropy per unit mass.
As can be seen from Ref.~\cite{landau}, the irreversible
fluxes of $c_{\rm a}$ and ${\check s}$ are respectively given by 
the quotient of the first entry in Eq.~(\ref{eqn:linear2}) divided by
$2\rho$ and that of the second divided by $\rho T$, whereas the
{conjugate} thermodynamic forces are {respectively given by}
$-2\rho({\bar \delta}\mu_-)/T$ and $-\rho ({\bar \delta }T)/T$.
After the division,
the second term on the RHS of the second entry becomes equal to the product of
 $-\rho(\nabla T)/T$ multiplied by ${\tilde \gamma}/\rho^2$, which is one of the Onsager coefficients.
Similarly, we can obtain the {other Onsager coefficients}. 
Comparing Eq.~(\ref{eqn:linear1}) with (\ref{eqn:linear2}), we obtain
\begin{equation}
\Lambda=4{\tilde \alpha}\ ,
\quad \kappa= 2T\left(2\mu_-{\tilde\alpha}-T{\tilde\beta}\right)\ ,
\quad {\rm and}\quad \lambda=4\mu_-T \left(\mu_-{\tilde\alpha}-T{\tilde\beta}\right) +T^2{\tilde\gamma}
\ .\label{eqn:kankei}\end{equation}
{In Refs.~\cite{mist, jetp, luet},
the singular parts of the LHS's above, indicated by
the subscript $_{\rm sing}$, are shown to satisfy}
\begin{equation}
\frac{{\tilde \beta}_{\rm sing}}{2{\tilde \alpha}_{\rm sing}}=-\left.\frac{\partial c_{\rm a}}{\partial T}\right)_{P, \mu_-}
\left.\frac{\partial \mu_-}{\partial c_{\rm a}}\right)_{T, P}=-{\bar s}_-\label{eqn:abgratio1}\ ,
\end{equation}
whose second equality comes from Eq.~(\ref{eqn:partials}), and
\begin{equation}
\frac{{\tilde \gamma}_{\rm sing}}{4{\tilde \alpha}_{\rm sing}}=\frac{\rho T}{\chi}\left.\frac{\partial {\check s}}{\partial T}\right)_{P, \mu_-}
\approx T{\bar s}^2_-\ . \label{eqn:abgratio2}\end{equation}
The approximate equality of  Eq.~(\ref{eqn:abgratio2}) is explained in the next paragraph.
The background parts of ${\tilde \alpha}$
and ${\tilde \beta}$ are negligible in the critical regime 
\cite{mirz,iwan,gigl,folkmos2}.  Because of the singular properties, 
the Ludwig-Soret effect has universal properties in a near-critical binary fluid mixture
\cite{mist,gigl,luet,ryzh,kohl}.
For ${\tilde \gamma}$, neglecting the background part
and adopting the approximate equality, we obtain Eq.~(\ref{eqn:kaplam}), which leads to neglect of
the thermal conductivity not exhibiting the critical enhancement \cite{luet}.
\\


\noindent
The partial derivative in Eq.~(\ref{eqn:abgratio2}) equals
\begin{equation}
\left.\frac{\partial {\check s}}{\partial T}\right)_{P, c_{\rm a}}+\left.\frac{\partial {\check s}}{\partial c_{\rm a}}\right)_{T, P}
\left.\frac{\partial c_{\rm a}}{\partial T}\right)_{P, \mu_-}
=\frac{c_P}{\rho T}+2{\bar s}_-
\left.\frac{\partial c_{\rm a}}{\partial T}\right)_{P, \mu_-}
\ ,\label{eqn:sbarT}\end{equation}
where $c_P$ denotes the isobaric specific heat under constant $c_{\rm a}$.
The equality between the second terms on both sides above comes from a Maxwell relation {and 
Eq.~(\ref{eqn:partials})}.
The second partial derivative in Eq.~(\ref{eqn:abgratio1})
can be rewritten using Eq.~(34) of Ref.~\cite{pipe}.  As a result, the last partial derivative 
in Eq.~(\ref{eqn:sbarT}) equals ${\bar s}_- \chi/(2\rho^2{\bar v}_+)$.
As mentioned in Appendix \ref{sec:fbulk}, {$c_P$ diverges more weakly
than $\chi$} in the critical regime,
where the approximate equality in Eq.~(\ref{eqn:abgratio2}) is valid. 
Without the above-mentioned approximations for ${\tilde \gamma}$,
the area integral of $T^2$ multiplied by the sum of the thermal conductivity and $\Lambda c_P/\chi$, evaluated at $\varepsilon=0$,
over $S_{\rm tube}$ should be added to the RHS of Eq.~(\ref{eqn:prel33}).
\\

\noindent
An alternative explanation is as follows. 
In the mode-coupling theory, the singular {part of an Onsager coefficient is} calculated in terms of
the time-integral of the two-time correlation function of the reversible
fluxes, as mentioned in  Section 6.5 of Ref.~\cite{onukibook}. 
The Onsager coefficients {mentioned above Eq.~(\ref{eqn:kankei})} are found to be
${\tilde \gamma}/\rho^2$, $-{\tilde \beta}T/\rho^2$, and ${\tilde \alpha} T/\rho^2$, 
and their singular parts
are linked with the autocorrelation of $({\bar \delta} {\check s}) {\v v}$,
the crosscorrelation of $({\bar \delta} {\check s}){\v v}$ and $({\bar \delta} c_{\rm a}){\v v}$,
and the autocorrelation of $({\bar \delta} c_{\rm a}){\v v}$, respectively.  
This means that, in the critical regime, the ratio of
${\tilde \gamma}_{\rm sing}$ to $-{\tilde \beta}_{\rm sing}T$ and 
that of $-{\tilde \beta}_{\rm sing}T$  to ${\tilde \alpha}_{\rm sing} T$
are given by $\partial {\check s}/(\partial c_{\rm a})$ with $T$ and $P$
fixed, which is consistent with Eqs.~(\ref{eqn:abgratio1}) and (\ref{eqn:abgratio2}). 
\end{section}

\begin{section}{{Internal-energy fluctuations} and approximate incompressibility \label{sec:fbulk} }
{Here, we discuss some backgrounds of Eq.~(\ref{eqn:efu}).} 
In the bare model,
without the contribution from $\partial V_{\rm tot}$ taken into account,
we define the effective Hamiltonian ${\cal H}$ so that the
equilibrium probability density functional (EPDF) of $\rho$ and $\psi$ is proportional to
$e^{-{\cal H}}$.  As mentioned in the text, ${\cal H}$ includes the $\psi^4$ model,
\begin{equation}
\int_{V_{\rm tot}} d{\v r}\ \left[ \frac{1}{2}A_0\tau \psi^2 +\frac{\lambda_0}{4!} \psi^4 + \frac{a_0^2}{2} \left|\nabla \psi\right|^2
\right]
\ ,\label{eqn:bare}\end{equation}
where $\lambda_0(>0)$ and $a_0$ are constants, and 
the integrand above becomes a part of Eq.~(\ref{eqn:efu}) divided by $k_{\rm B}T$ after coarse-grained.
The value of $T_{\rm c}$ in the definition of $\tau$ depends on the stage of the
coarse-graining. Writing $m$ for $(u-u_{\rm c})/(k_{\rm B}T_{\rm c})$,
we can also consider the EPDF of $\rho$, $\psi$, and $m$.
We define ${\cal H}_{\rm s}$ so that this EPDF is proportional to $e^{-{\cal H}_{\rm s}}$ in the bare model.
Integrating out $m$ from this EPDF should yield the
EPDF of $\rho$ and $\psi$.  In other words, the latter's Legendre transform is the former, and vice versa.
{Thus, owing to a term $A_0\tau\psi^2/2$ in ${\cal H}$},  ${\cal H}_{\rm s}$  has
a  term proportional to $m\psi^2$ and  ${\cal H}$ has
a term proportional to $\tau^2$ \cite{halhoma, onukibook}.  
We define $C_0$ so that this term equals $-C_0\tau^2/(2k_{\rm B})$,
and the variance of $m$ is proportional to $C_0$.
Coarse-graining the $\psi^4$ model
and imposing the self-consistent condition for off-critical compositions
set up the RLFT \cite{rlft}.  We can also set it up by
coarse-graining ${\cal H}_{\rm s}$, imposing a self-consistent condition, and
integrating out $m$.   In this procedure, $C_0$ becomes $C$ of Eq.~(\ref{eqn:efu})  because 
fluctuations of $m$ are affected by those of $\psi$ via their coupling term
\cite{halhoma, onukibook}.  Instead of using Eq.~(\ref{eqn:efu}) as it is,
we can calculate {${\hat\sigma}^{({\rm th})}$} by evaluating
the dependence of $C$ on $\xi$ locally. 
Although data are not shown, the resultant changes {from the results of Fig.~\ref{fig:inte}} are
negligibly small.  
\\

\noindent
The isochoric specific heat $c_V$, given by $-T\partial^2 f_{\rm bulk}/(\partial T^2)$ with
$\rho$ and $\varphi$ being fixed,  
remains finite at the critical point, although it appears to diverge in the regime
accessible to usual experiments \cite{anisim72, onukibook}.
Linked with the fluctuations of $m$, {the isobaric specific heat} $c_{P}$
becomes proportional to $C\ (\propto |\tau|^{-\alpha}$ with $\alpha=0.11)$
 in the critical regime with $c_{\rm a}$ {being the value at the critical point} \cite{anisim95, onukibook}.  
The same power-law dependence is shared by
the isothermal compressibility $\kappa_{T}$ and the thermal expansion coefficient $\alpha_{P}$, which 
are given by
\begin{equation}
{\kappa_{T}}=\frac{1}{\rho}\left.\frac{\partial \rho}{\partial P}\right)_{T, c_{\rm a}}\quad{\rm and}\quad
{\alpha_{P}}=-\frac{1}{\rho}\left.\frac{\partial \rho}{\partial T}\right)_{{P}, c_{\rm a}}=
{\kappa_{T}}\left.\frac{\partial P}{\partial T}\right)_{\rho, c_{\rm a}}
\ .\label{eqn:kapalp}\end{equation}
These quantities are related with each other via
$\kappa_{T}\left( c_{P}-c_{V} \right)= \alpha_{P}^2 T$.
The singularity of $\kappa_{T}$ is generated by
 coupling between $\rho-\rho_{\rm c}$ and $\psi^2$ in the $\rho$-dependent part
in the EDPF\@. We neglect this coupling in Eq.~(\ref{eqn:efu}),
considering that the singularity is not accessible to
usual experiments \cite{clerke, onukibook}.  
Observed values of $\kappa_T$ and $\alpha_P$ are typically 
$10^{-9}\ $Pa$^{-1}$ and  $10^{-3}\ $K$^{-1}$, respectively, near the critical point \cite{clerke, pous}.
\\

\noindent
{In this paragraph, we show that the last partial derivative of Eq.~(\ref{eqn:diss1}) is finite.
In the region mentioned at the beginning of Appendix \ref{sec:diss},} 
we simply write $f$ for $f_{\rm bulk}$ and refer to
its derivatives by adding subscripts. For example, 
$f_{\rho\varphi}$ represents  $\partial^2 f_{\rm bulk}/(\partial \rho \partial \varphi)$, which vanishes
because Eq.~(\ref{eqn:efu}) is assumed.  We have 
\begin{equation}
\delta P=\rho\left(f_{\rho\rho}\delta\rho + f_{\rho T}\delta T\right)+\varphi\left(
{f_{\varphi\varphi}\delta \varphi} +f_{\varphi T}\delta T\right)
+s\delta T\label{eqn:deltaP2}
\end{equation}
owing to Eq.~(\ref{eqn:deltaP}).  Here, {unlike in the text, $\delta$} indicates an infinitesimal change.
Using Eq.~(\ref{eqn:deltaP2}) and 
$\varphi=\rho(2c_{\rm a}-1)$, we find $\delta \varphi$  ($\delta \rho$)
equal to $\varphi$ ($\rho$) multiplied by
\begin{equation}
\delta P \frac{1}{\rho^2f_{\rho\rho}+\varphi^2f_{\varphi\varphi}}\label{eqn:newkapT}
\end{equation} when $T$ and $c_{\rm a}$ are fixed.
The fraction is found to equal $\kappa_{T}$ because of the first entry of Eq.~(\ref{eqn:kapalp}) and Eq.~(\ref{eqn:deltaP2}).
Considering that the sum in the first (second) parentheses 
of Eq.~(\ref{eqn:deltaP2}) equals $\delta\mu_+$ ($\delta\mu_-)$,
we use  Eq.~(\ref{eqn:newkapT}) to find that the first entries of
Eqs.~(\ref{eqn:partials}) and (\ref{eqn:rhovv}) give  ${\bar v}_-=\varphi f_{\varphi\varphi}\kappa_T$
and ${\bar v}_+=\rho f_{\rho\rho}\kappa_T$.  Thus, we use Eq.~(\ref{eqn:deltaP2})
to find that the last partial derivative in Eq.~(\ref{eqn:diss1}) equals $-{\bar v}_-/{\bar v}_+$,
which result holds if $f_{\rho\varphi}$ does not vanish.
The limit of this fraction obtained as the critical point is approached {can 
be} written in terms of $\rho_{\rm c}$ and $\varphi_{\rm c}$ and is finite,
which supports the description below Eq.~(\ref{eqn:diss1}). \\

\noindent
Finally, we consider validity of $\rho^{({\rm ref})}\approx \rho_{\rm c}$; $\rho^{({\rm ref})}$ appears
in Eq.~(\ref{eqn:PL1}) and is involved in returning the dimension to ${\hat G}^{({\rm th})}$.
For definiteness, we here write $\tau^{({\rm ref})}$ for the value of $\tau$ in the reference state.
At the state we reach by changing $\tau$ from zero to $\tau^{({\rm ref})}$ with $P$ and $c_{\rm a}$ being fixed, 
how $\rho$ changes from $\rho_{\rm c}$ can be approximately 
calculated from the regular part of $\alpha_P$ \cite{clerke}.
The difference between the value of $\rho$ {at this state} and $\rho^{({\rm ref})}$ 
can be calculated using the last partial derivative in Eq.~(\ref{eqn:diss1}).
Thus, we can calculate the difference $\rho^{({\rm ref})}-\rho_{\rm c}$, and find that
the difference divided by $\rho_{\rm c}$ is smaller than  {$10^{-3}$} 
for ${\tau^{({\rm ref})}}=1/300$ in magnitude. 
Thus, we can use $\rho^{({\rm ref})}\approx \rho_{\rm c}$.
\end{section}

\begin{section}{Approximation 
in the derivation of {Eq.~(\ref{eqn:energy-density})}\label{sec:detail}}
{By} using Eq.~(\ref{eqn:deltaP2}) to calculate the second entry of
Eq.~(\ref{eqn:partials}), we obtain ${\bar s}_-= \varphi f_{\varphi\varphi}\alpha_P-f_{\varphi T}$.
The term $\mu_-^{({\rm ref})}\varphi$ is included in $f_-$, as shown in Eq.~(\ref{eqn:prefmm}).
Its contribution to the second entry of
Eq.~(\ref{eqn:sandu}) is 
\begin{equation}
-T^2\left.\frac{\partial}{\partial T}\right)_{\rho, \varphi} \frac{\mu_-^{({\rm ref})}\varphi}{{T}} =
\mu_-^{({\rm ref})}{\varphi
-T\varphi}\left.\frac{\partial}{\partial T}\right)_{\rho, \varphi} {\mu_-^{({\rm ref})}}
\ .\label{eqn:murefcont}\end{equation}
The value of the last partial derivative above equals that of $f_{\varphi T}$ {in the reference state} owing to $f_{\rho\varphi}=0$.
Thus, because of Eq.~(\ref{eqn:barHn}), Eq.~(\ref{eqn:murefcont})
consists of ${\bar H}_-^{({\rm ref})}\varphi$ and the other term.
This term gives an extra term  
\begin{equation}
{\frac{T_*\tau_*}{\mu_*}} \left(\varphi f_{\varphi\varphi}\alpha_{\rm P}\right)^{({\rm ref})}{\hat\psi}^{(0)} 
\label{eqn:include}\end{equation}
to Eq.~(\ref{eqn:thermal-force-density0}), 
where the superscript $^{({\rm ref})}$ implies that the product in the parentheses
should be evaluated in the reference state.
We can use Eq.~(\ref{eqn:fprpr}) to evaluate $f_{\varphi\varphi}$ and find Eq.~(\ref{eqn:include})
to be smaller than $30|\tau|^\gamma |{\hat\psi}^{(0)}|$ in magnitude.
This magnitude is found to be 
much smaller than the corresponding magnitude given by the circles in Fig.~\ref{fig:inte},
with the aid of values of ${\hat\psi}^{(0)}$ in Fig.~\ref{fig:profile}.
{Thus, Eq.~(\ref{eqn:include})} is negligible
in deriving Eq.~(\ref{eqn:energy-density}).
\end{section}

\begin{section}{Thermoosmotic flow far from a flat surface\label{sec:gauss}}
For a mixture occupying a semi-infinite space bounded by a flat wall surface,
we consider imposing a temperature gradient along the $z$ axis, which is parallel to the wall surface.
{The equilibrium profile
$\psi^{(0)}$ is regarded as a function of the distance from the surface, denoted by $X$,
and is assumed to approach zero} as $X\to\infty$.
The velocity field can be calculated similarly to Eq.~(\ref{eqn:nodimv0}).
Assuming that $\eta_{\rm s}$ to be $\eta_*$ homogeneously and {using Eq.~(\ref{eqn:thermalapp2}), 
we find that} the $z$ component of the mixture velocity far from the surface, or the slip velocity,
 is approximately given by
\begin{equation}
\frac{k_{\rm B}C_1{|\tau|}}{2\eta_*\xi_0^2 \tau}\int_0^XdX_1\int_{X_1}^\infty dX\ \omega^{\gamma-1}\left[\psi^{(0)}(X)\right]^2
\label{eqn:flat} 
\end{equation}
multiplied by the $z$ component of the temperature gradient. 
\\

\noindent
When the second term is much smaller than the first term on the RHS of Eq.~(\ref{eqn:omega2}),
${\hat \omega}\approx {|{\hat \tau}|}$ holds and {the second term on the RHS} 
of Eq.~(\ref{eqn:prefmm0}) is negligible.  We further approximate $M_-$
to be $k_{\rm B}T_{\rm c} C_1$ to obtain {the free-energy density
in the Gaussian model, where $\xi$ becomes homogeneous} 
and $\psi^{(0)}(X)$ equals $h\xi e^{-X/\xi}/M_-$ \cite{undul, EPJ}. 
Substituting this into Eq.~(\ref{eqn:flat}) with $\omega=|\tau|$ and $h=0.1\ $cm$^3/$s$^2$, 
we find that the slip velocity {in terms of ${\hat v}_z^{({\rm th})}$ for a mixture of LW (NEMP)  is $-0.082\ (0.12)$ 
at $|\tau|=3.2\times 10^{-3}$, and  $-0.017\ (0.025)$ 
at $|\tau|=6.4\times 10^{-3}$.  Here, {for the critical exponents,
we use the values mentioned in Section \ref{sec:free}, not the values in the Gaussian model}.
These values of the slip velocity are comparable with
the corresponding values calculated in the same procedure as used for Fig.~\ref{fig:vel}, which 
are $-0.042\ (0.061)$ at $|\tau|=3.2\times 10^{-3}$ and $-0.014\ (0.021)$ at $|\tau|=6.4\times 10^{-3}$.}\\

\noindent
The equilibrium profile $\psi^{(0)}(X)$
becomes universal {in the adsorption layer} as the critical point is approached 
beyond the regime of the Gaussian model \cite{fisher-degennes,RJ,diehl97}. 
As is done for the diffusioosmosis in Appendix D of Ref.~\cite{pipe}
and in Section VC of Ref.~\cite{diffdrop}, 
{we use the universal profile $\psi^{(0)}(X)\propto X^{-\beta/\nu}$ 
in Eq.~(\ref{eqn:flat})}.  With the aid of Eq.~(\ref{eqn:omega2}), we find that
the slip velocity in thermoosmosis becomes proportional to {$|\tau|^{\nu-1}$} as $\tau$ approaches zero.  \\
\end{section}


\begin{thebibliography}{99}
\bibitem{breg} A. P. Bregulla, A. W\"{u}rger, K. G\"{u}nther, M. Mertig, and F. Cichos, ``Thermo-osmotic flow in thin films," Phys.~Rev.~Lett. 
{\bf 116}, 188303 (2016).
{\bibitem{lee} C. Lee, C. Cottin-Bizonne, A.-L. Biance, P. Joseph, L. Bocquet, and C. Ybert,
``Osmotic flow through fully permeable nanochannels," Phys.~Rev.~Lett.~{\bf 112}, 244501 (2014).}
\bibitem{shin} S. Shin, 
``Diffusiophoretic separation of colloids in microfluidic flows," Phys.~Fluids {\bf 32}, 101302 (2020). 
\bibitem{shakib} S. Shakib, B. Rogez, S. Khadir, J. Polleux, A. W{\"u}rger, and G. Baffou,
 ``Microscale thermophoresis in liquids induced by plasmonic heating and characterized by phase 
and fluorescence microscope," J. Phys.~Chem.~C {\bf 125}, 21533--21542 (2021).
\bibitem{chen} W. Q. Chen, M. Sedighi, and A. P. Jivkov, 
``Thermo-osmosis in hydrophilic nanochannels: mechanism and size effect," Nanoscale {\bf 13}, 1696--1716 (2021).
\bibitem{ganti} R. Ganti, Y. Liu, and D. Frenkel, "Molecular simulation of thermo-osmotic slip," 
Phy.~Rev.~Lett.~{\bf 119} 038002 (2017).
\bibitem{piazza}
R. Piazza and A. Parola, ``Thermophoresis in colloidal suspensions,"
J.~Phys.:Condens.~Matter {\bf 20}, 153102 (2008).
{\bibitem{Wurger} A. W\"{u}rger,
``Thermal non-equilibrium transport in colloids",
Rep.~Prog.~Phys.~{\bf 73}, 126601 (2010).}
\bibitem{marbach} S. Marbach and L. Bocquet, ``Osmosis, from molecular insights to large-scale applications," 
Chem.~Soc.~Rev.~{\bf 48}, 3102-3144 (2019).
{\bibitem{mang} E. Mangaud and B. Rotenberg, ``Sampling mobility profiles of confined
fluids with equilibrium molecular dynamics simulations," J.~Chem.~Phys.~{\bf 153}, 044125 (2020).}
\bibitem{frenkel} S. Ram{\' i}rez-Hinestrosa and D. Frenkel, ``Challenges in modelling diffusiophoretic transport,"
Eur.~Phys.~J. B {\bf 94}, 199 (2021).
\bibitem{derja} B. V. Derjaguin, N. Churaev, and V. Muller, {\it Surface Forces\/}
(Springer Science+Business Media, LLC, Berlin, 1987).
\bibitem{derja2} B. V. Derjaguin,
``Some results from 50 years' research on surface forces," In {\it Surface Forces and Surfactant Systems,
Progress in Colloid \& Polymer Science 74\/} (Steinkopff, Dresden, 1987) 17--30.
\bibitem{anders} J. L. Anderson, ``Colloid transport by interfacial forces," Ann.~Rev.~Fluid Mech.~{\bf 21}, 61--99 (1989).
\bibitem{derja3} B. V. Derjaguin and G. P. Sidorenkov, ``On thermo-osmosis of
liquid in porous glass," CR Acad.~Sci.~URSS
{\bf 32}, 622--626 (1941). 
{\bibitem{derja4}
B. V. Derjaguin, G. P. Sidorenkov, E. A. Zubashchenkov, and E. V. Kiseleva E V, ``Kinetic phenomena in boundary
films of liquids," Kolloidn.~Zh.~{\bf 9}, 335–347 (1947).
\bibitem{derja5} B. V. Derjaguin, S. S. Dukhin, and M. M. Koptelova,
``Capillary osmosis through porous partitions and properties of
boundary layers of solutions," J. Colloid.~Interface Sci.~{\bf 38}, 584--595 (1972). }
\bibitem{fu} L. Fu, S. Merabia, and L. Joly, ``What controls thermo-osmosis? Molecular simulations show the critical role of 
interfacial hydrodynamics," Phys.~Rev.~Lett.~{\bf 119}, 214501 (2017).
\bibitem{Anzini} P. Anzini, G. M. Colombo, Z. Filiberti,  A. Parola, ``Thermal forces from a microscopic perspective,"
Phys.~Rev.~Lett.~{\bf 123}, 028002 (2019).
\bibitem{pipe}
S. Yabunaka and Y. Fujitani, ``Isothermal transport of a near-critical binary fluid mixture through a capillary tube with the preferential adsorption,"
Phys.~Fluids {\bf 34}, 052012 (2022).
\bibitem{beysens1982}
D. Beysens and S. Leibler, "Observation of an
  anomalous adsorption in a critical binary mixture," J. Physique Lett.~{\bf 43}, 133--136  (1982).
{\bibitem{schol} M. Schlossman, X-L. Wu, and C. Franck,
``Order-parameter profile at long distances in an adsorbed binary liquid mixture near criticality,"
Phys.~Rev.~B {\bf 31}, 1478--1485 (1985).
\bibitem{binder}
M. N. Binder, {\it Phase Transitions and Critical
  Phenomena VIIIV\/}, Critical behavior at surfaces.  (Academic,
  London, 1983).
\bibitem{diehl86}
H. W. Diehl, {\it Phase Transition and Critical Phenomena
  X\/}, Field theoretical approach to critical behavior at surfaces.
  (Academic, London, 1986).}
\bibitem{diehl97}
H. W. Diehl, "The theory of boundary critical
  phenomena,"  Int.~J. Mod.~Phys.~B  {\bf 11}, 3503--3523 (1997).
{\bibitem{law} B. M. Law, ``Wetting, adsorption, and surface critical phenomena," Prog.~Surf.~Sci. {\bf 66}, 159-216 (2001).
\bibitem{yabufuji} S. Yabunaka and Y. Fujitani,
``Drag coefficient of a rigid spherical particle in a near-critical binary fluid mixture, beyond the regime of the Gaussian model,"
J. Fluid Mech.~{\bf 886} A2 (2020).}
\bibitem{kawasaki}  K. Kawasaki, ``Kinetic equations and time correlation functions of critical fluctuations," Ann.~Phys.~(N.Y.) {\bf 61}, 1 (1970)
\bibitem{sengers} J. V. Sengers, ``Transport properties near critical points," Int. J. Thermophys. {\bf 6}, 203--232 (1985).
\bibitem{sighalhoh}
E. D. Siggia, P. C. Hohenberg, and B. I. Halperin, 
``Renormalization-group treatment of the critical dynamics of the binary-fluid and gas-liquid transitions,"  Phys.~Rev.~B  {\bf 13},
2110--2123 (1976).
\bibitem{onukibook}  A. Onuki, {\it Phase Transition Dynamics\/}
  (Cambridge University Press, Cambridge, 2002), Sections 2.3, 3.1, 4.3, and 6.5.  
\bibitem{folkmos} R. Folk
and G. Moser, ``Critical dynamics: a field-theoretical approach," J. Phys.~A: Math.~Gen.~{\bf 39}, R207--R313 (2006).
\bibitem{fisher-degennes} M. E. Fisher and P. G. de Gennes, 
"Ph\'{e}nom\`{e}nes aux parois dans un m\'{e}lange binaire critique," C. R. Acad. Sci. Paris B~{\bf  287}, 207 (1978).
\bibitem{RJ} J. Rudnick and D. Jasnow, ``Order-parameter profile in semi-infinite systems at criticality,"
Phys.~Rev.~Lett.~{\bf 48}, 1059 (1982).
\bibitem{diffphore} Y. Fujitani, ``Diffusiophoresis in a near-critical binary fluid mixture,"
Phys.~Fluids {\bf 34}, 041701 (2022).
\bibitem{diffdrop} Y. Fujitani, ``Effects of the preferential adsorption 
in a near-critical binary fluid mixture on dynamics of a droplet,"
Phys.~Fluids {\bf 34}, 092007 (2022).
\bibitem{dvw} A. Onuki, ``Dynamic van der Waals theory," Phys.~Rev.~E {\bf 75}, 036304 (2007).
\bibitem{gonn} G. Gonnella, A. Lamura, and A. Piscitelli, ``Dynamics of binary fluid mixtures in inhomogeneous temperatures,"
J. Phys.~A: Math.~Theor.~{\bf 41} 105001 (2008). 
\bibitem{rlft} R. Okamoto and A. Onuki, 
``Casimir amplitude and capillary condensation of near-critical binary fluids between parallel plates: renormalized local functional theory,"
  J. Chem.~Phys.~{\bf 136}, 114704 (2012).
{\bibitem{complett}
S. Yabunaka and Y. Fujitani, ``Universal direction in thermoosmosis of a near-critical binary fluid mixture," submitted.}
 \bibitem{gromaz} S. R. de Groot and G. Mazur, {\it Non-equilibrium thermodynamics\/}, (Dover, New York, 1984).
Chapters XI and XV.
\bibitem{engineer} S. Kjelstrup, D. Bedeaux, E. Johannessen, and J. Gross, {\it Non-Equilibrium Thermodynamics for Engineers\/}
  (World Scientific, New Jersy, 2017).
{\bibitem{halhohsig} B. I. Halperin,  P. C. Hohenberg, and E. D. Siggia, 
``Renormalization-group calculations of divergent transport coefficients at critical point,"
Phys.~Rev.~Lett.~{\bf 32}, 1289 (1974).
\bibitem{ohta} T. Ohta, ``Selfconsistent calculation of dynamic critical exponents for classical liquid,"
Prog.~Theor.~Phys.~{\bf 54}, 1566 (1975).}
\bibitem{mirz} S. Z. Mirzaev, R. Behrends, T. Heimburg, J. Haller, and U. Kaatze, 
``Critical behavior of 2,6-dimethylpyridine-water: Measurements of specific heat, dynamic light scattering, and shear viscosity," 
J. Chem. Phys. {\bf 124} 144517 (2006).
\bibitem{mist} L. Mistura, ``Transport coefficients near a critical point in multicomponent fluid system," Nuovo Cimento B {\bf 12}, 35--42.
\bibitem{sciam} J. S. Walker and C. A. Vause, `` Reappearing phases,"
Sci.~Am.~{\bf 256}, 98-100 (1987).
\bibitem{kaji} M. Toda, S. Kajimoto, S. Toyouchi, T. Kawakatsu, Y. Akama, M. Kotani, and H. Fukumura,
``Phase behavior of a binary fluid mixture of quadrupolar molecules," Phys.~Rev.~E {\bf 94}, 052601 (2016).
\bibitem{tsori} Z. Chernia and  Y. Tsori, ``Hydrogen bonding of dimethylpyridine
clusters in water: correlation between the
lower consolute solution temperature and
electron interaction energy," J. Chem.~Phys.~{\bf 152}, 204304 (2020).
\bibitem{peli}
A. Pelisetto and E. Vicari, ``Critical phenomena
  and renormalization-group theory,"  Phys.~Rep.~{\bf 368},  549 (2002).
\bibitem{halhoma} B. I. Halperin, P. C. Hohenberg, and S. Ma, ``Renormalization-group methods for critical dynamics: I. recursion relations
and effects of energy conservation," Phys.~Rev.~B {\bf 10},  139--153 (1973).
\bibitem{bergmold}
R. F. Berg and M. R. Moldover, 
``Critical exponent for the viscosity of four binary liquids,"  J. Chem.~Phys.
{\bf 89}, 3694--3704 (1989). 
\bibitem{bergmold2} R. F. Berg and M. R. Moldover, 
``Critical exponent for  viscosity,"  Phys.~Rev.~A
{\bf 42}, 7183--7187 (1990).
\bibitem{edison} T. A. Edison, M. A. Anisimov, and J. V. Sengers, ``Critical scaling laws and an excess Gibbs energy model,"
Fluid Phase Equilibria {\bf 150}--{\bf 151}, 429--438 (1998).
\bibitem{vanthof} A. van't Hof, M. Laura Japas, and C. J. Peters, 
``Description of liquid-liquid equilibria including the critical region with the
crossover-NRTL model," Fluid Phase Equilibria {\bf 192}, 27--48 (2001).
\bibitem{olaya} M. M. Olaya, P. Carbonell-Hermida, M. Trives, J. A. Labarta, and A. Marcilla,
``Liquid-liquid equilibrium data correlation using NRTL model for different types of binary systems: upper critical solution temperature , lower
critical solution temperature, and closed miscibility loops,"
 Ind.~Eng.~Chem.~Res.~{\bf 59}, 8469--8479 (2020).
\bibitem{iwan} I. Iwanowski, K. Leluk, M. Rudowski, and U. Kaatze, 
``Critical dynamics of the binary system nitroethane/3-methylpentane: Relaxation rate and
scaling function," J. Phys.~Chem.~A {\bf 110}, 4313--4319 (2006).
\bibitem{to} K. To,  ``Coexixtence curve exponent of a binary mixture with a high molecular weight polymer,"
Phys.~Rev.~E. {\bf 63}, 026108 (2001). 
\bibitem{wims} A. M. Wims, D. Mcintyre, and F. Hynne,
 ``Coexistence curve for 3-methylpentane-nitroethane near the critical point," J. Chem.~Phys.~{\bf 50} 616 (1969).
\bibitem{bhatt} J. K. Bhattacharjee, R. A. Ferrell, R. S. Basu, and J. V. Sengers, 
``Crossover function for the critical viscosity of a classical fluid," Phys. Rev. A, {\bf 24}, 1469 (1981).
\bibitem{tsai}  B. C. Tsai and D. McIntyre, "Shear viscosity of nitroethane-3-methylpentane in the critical region,"
J.~Chem.~Phys.~{\bf 60}, 937 (1974).
\bibitem{stein} A. Stein, S. J. Davidson, J. C. Allegra, and G. F. Allen, 
``Tracer Diffusion and Shear Viscosity for the System 2,6-Lutidine-Water near the
Lower Critical Point," J. Chem. Phys. {\bf 56} 6164 (1972).
\bibitem{gratt} C. A. Grattoni, R. A. Dawe, C. Y. Seah, and J. D. Gray, 
``Lower Critical Solution Coexistence Curve and Physical Properties
(Density, Viscosity, Surface Tension, and Interfacial Tension) of
2,6-Lutidine $+$ Water," Chem. Eng. Data, {\bf 38}, 516--519 (1993).
\bibitem{leis} H. M. Leister, J. C. Allegra, and G. F. Allen, 
``Tracer diffusion and shear viscosity in the liquid-liquid critical region," J. Chem.~Phys.~{\bf 51} 3701 (1969).
\bibitem{Braun} D. Braun, and A. Libchaber, ``Trapping of DNA by thermophoretic depletion and convection," Phys.~Rev.~Lett.~{\bf 89}, 
188103 (2001).
\bibitem{Jiang}H. Jiang, H. Wada, N. Yoshinaga, and M. Sano,  ``Manipulation of colloids by a nonequilibrium depletion force in a temperature gradient",
Phys.~Rev.~Lett.~{\bf 102}, 208301 (2009).
\bibitem{maeda} Y. T. Maeda, A. Buguin, and A. Libchaber,  ``Thermal separation: interplay between the Soret effect and entropic force gradient,"
Phys.~Rev.~Lett. ~{\bf 107}, 038301 (2011).
{\bibitem{omari}
R. A. Omari, C. A. Grabowski, and A.  Mukhopadhyay, ``Effect of surface curvature on
critical adsorption," Phys.~Rev.~Lett.~{\bf 103}, 225705 (2009).
\bibitem{beys2019}
D. Beysens, ``Brownian motion in strongly fluctuating liquid," Thermodyn. Interfaces Fluid Mech.
{\bf 3}, 1–-8 (2019).}
\bibitem{wegner} F. J. Wegner, ``Corrections to scaling laws," Phys.~Rev.~B {\bf 5}, 4529--4536 (1972).
\bibitem{CAS} Z. Y. Chen, P. C. Albright, and J. V. Sengers, ``Crossover from
singular critical to regular classical thermodynamic behavior of fluids," Phys.~Rev.~A {\bf 41}, 3161 (1990).
\bibitem{CATS}  Z. Y. Chen, P. C. Albright, S. Tang, and J. V. Sengers, ``Global thermodynamic
behavior of fluids in the critical region," Phys.~Rev.~A {\bf 42}, 4470 (1990).
\bibitem{folkmos2} R. Folk
and G. Moser, ``Critical dynamics in mixtures,"   Phys.~Rev.~E {\bf 58}, 6246--6274 (1998).
\bibitem{degennes}  De Gennes, P. G. and Prost, J. (1993). {\it The physics of liquid crystals \/} (Oxford Univ.~Press, Oxford, 1993).
\bibitem{physica}  Y. Fujitani, ``Dynamics of the lipid-bilayer membrane taking a vesicle shape," Physica A {\bf 203},  214--242 (1994).
[Erratum: {\it ibid}  {\bf 237}, (1997) 346--347].
\bibitem{effvis}
Y. Fujitani, ``Effective viscosity of a near-critical
  binary fluid mixture with colloidal particles dispersed dilutely under weak
  shear,"  J.~Phys.~Soc.~Jpn.~{\bf 83},  084401 (2014).
\bibitem{luet} {J. Luettmer-Strathmann}, ``Thermal diffusion in the critical region," 
In {\it Thermal nonequilibrium phenomena in Fluid Mixtures
(Letcture Notes in Physics 584)\/}
(Springer, Berlin, 2002), 24--37.
\bibitem{landau}
L. D. Landau and E. M. Lifshitz, {\it Fluid Mechanics, 2nd Ed.\/}
(Elsevier, Amsterdam, 1987), \S 58.
\bibitem{jetp} M. A. Anisimov, A. V. Voronel, and E. E. Gorodetskii, 
``Isomorphism of critical phenomena," Sov.~Phys.~JETP {\bf 33}, 605 (1971). 
\bibitem{gigl} M. Giglio and A. Vendramini, ``Thermal-diffusion mesurements near a consolute critical point," Phys.~Rev.~Lett.~{\bf 34} 561--564 (1975).
\bibitem{ryzh} I. I. Ryzhkov and S. V. Kozlova, ``Stationary and transient Soret separation in a binary mixture with a consolute critical point,"
Eur.~Phys.J. E {\bf 39}, 130 (2016).
\bibitem{kohl} W. K{\"o}hler and K. I. Mozorov, ``The Soret effect in liquid mixtures -- a review,''  J. Non-Equilib.~Thermodyn.~{\bf 41}, 151-197 (2016).
\bibitem{anisim72}  M. A. Anisimov, A. V. Voronel, and T. M. Ovodova, ``The behavior of thermodynamical quantities near the critical line of an 
incompressible liquid mixture", Sov.~Phys.~JETP {\bf 35}, 536--539 (1972).
\bibitem{anisim95} M. A. Anisimov, E. E. Gorodetskii, V. D. Kulikov, 
A. A. Povodyrev, J. V. Sengers, ``A general isomorphism approach to 
thermodynamic and transport properties of binary 
fluid mixtures near critical points,"  Physica {\bf A 220}, 277--324 (1995). 
\bibitem{clerke} E. A. Clerke and J. V. Sengers, ``Fast pressure quenches near the critical point of a binary liquid mixture,"
Physica {\bf 118A}, 360--370 (1983).
\bibitem{pous} F. Pousaneh, O. Edholm, and A. Macio{\l}ek, ``Molecular dynamics simulation of a binary
mixture near the lower critical point,"  J. Chem.~Phys.~{\bf 145}, 014501 (2016).
\bibitem{undul}  Y. Fujitani, ``Undulation amplitude of a fluid membrane
 in a near-critical binary fluid mixture calculated beyond the Gaussian model
 supposing weak preferential attraction," J. Phys.~Soc.~Jpn. {\bf 86}, 044602 (2017).
\bibitem{EPJ} Y. Fujitani, 
``Relaxation rate of the shape fluctuation of a fluid membrane immersed in a near-critical binary fluid mixture,"
 Eur.~Phys.~J. E {\bf 39}, 31 (2016). 
\end{thebibliography}
\end{document}